\documentclass[a4paper]{article}
\usepackage{cite}
\usepackage{graphicx}
\usepackage{amsmath}
\usepackage{amssymb,bm}
\usepackage{array}
\usepackage{subfigure}
\usepackage{url}
\usepackage{algorithm}
\usepackage{algorithmic}
\usepackage{multirow}
\usepackage[margin=20truemm]{geometry}

\def\myinter#1{\boldsymbol{[}{#1}\boldsymbol{)}}
\def\myInter#1{\boldsymbol{\biggl [}{#1}\boldsymbol{\biggr )}}

\makeatletter

\newcommand{\Rmnum}[1]{\expandafter\@slowromancap\romannumeral #1@}

\def\mytheory#1#2#3{\vspace{3mm} \hrule \vspace{1mm} \begin{#1}[#2] #3 \end{#1} \vspace{1mm} \hrule \vspace{3mm}}
\makeatother

\begin{document}
\title{General Form of Almost Instantaneous \\Fixed-to-Variable-Length Codes}

\author{Ryosuke~Sugiura,
        Yutaka~Kamamoto,
        and~Takehiro~Moriya}
\date{}
\maketitle

\begin{abstract}
A general class of the almost instantaneous fixed-to-variable-length (AIFV) codes is proposed, 
which contains every possible binary code we can make when allowing finite bits of decoding delay. 
The contribution of the paper lies in the following. 
(i) Introducing $N$-bit-delay AIFV codes, constructed by multiple code trees with higher flexibility than the conventional AIFV codes. 
(ii) Proving that the proposed codes can represent any uniquely-encodable and uniquely-decodable variable-to-variable length codes. 
(iii) Showing how to express codes as multiple code trees with minimum decoding delay. 
(iv) Formulating the constraints of decodability as the comparison of intervals in the real number line. 
The theoretical results in this paper are expected to be useful for further study on AIFV codes. 
\end{abstract}

%%%%%%%%%%%%%%%%%%%%%%%%%%%%%%%%%%%%%%%%%%%%%%%%%%%%%%%%%%%%%%%%%%%%%%%%%%%%%%%%
\section{Introduction}
\label{sec:intro}
For years, years, data compression techniques have greatly contributed to the development of many coding applications, such as audio and video codecs \cite{ref:MSSN, ref:hbdc,ref:audio_coding, ref:speech_coding}, which are now essential for our communication tools. 
Especially, lossless compression is one of the fundamental factors even for lossy situations \cite{ref:grcapp1, ref:tcx}.
In audio and video codecs, compression schemes are often required to encode given sequences of input signals with their distributions assumed using some models. 
There are two well-known approaches for compression in these cases: 
Huffman coding \cite{ref:huffman, ref:candc} gives us the minimum-redundancy codes among instantaneously decodable fixed-to-variable-length (FV) codes;
arithmetic coding \cite{ref:hbdc, ref:candc, ref:introDC} gives us variable-to-variable-length (VV) codes which are not necessarily minimum redundancy but asymptotically achieve entropy rates when the input sequence is long enough.  

Although Huffman coding guarantees its optimality, it shows lower compression efficiency compared to the arithmetic coding for many practical cases. 
This fact is mainly due to its constraint of instantaneous decodability, which strongly restricts the flexibility of the codeword construction: 
Instantaneous FV codes only allow the set of codewords that can be represented as a single code tree, 
with the input source symbols separately assigned to its leaves. 

Yamamoto ${\it et}$ ${\it al}$.~proposed a more flexible class of FV codes, 
the almost instantaneous FV (AIFV) codes \cite{ref:aifv1, ref:opt_aifv_dp, ref:opt_aifv_dp_fast}.
They loosen the constraints mentioned above by permitting, in binary code symbol cases, 
two bits of delay for decoding. 
This relaxation enables us to construct two code trees to represent codewords with more freedom for the source symbol assignment.  
This scheme is extended as AIFV-$m$ codes \cite{ref:aifv2, ref:opt_aifvm, ref:opt_aifvm_dp, ref:aifvm_red}, 
which permit $m$ bits of decoding delay to use $m$ code trees to represent codewords. 

AIFV-$m$ codes have more freedom for codeword construction than the instantaneous ones and can potentially outperform Huffman codes. 
However, we cannot say that AIFV-$m$ codes are fully using the advantage of decoding delay relaxation. 
Our previous works \cite{ref:xdgr,ref:xdg} revealed some types of practical AIFV codes 
that do not follow the rules for AIFV-$m$ codes. 
This paper aims to show what kind of code trees we can actually construct under a given decoding delay, 
which must be useful to make better use of the almost-instantaneous condition. 

The paper first prepares the basic ideas in Section \ref{sec:prepare}, reviewing the conventional AIFV codes and clarifying the general definition of decoding delay. 
Then, in Section \ref{sec:prop}, we define a broader class of AIFV codes proving its decodability and generality. 
Fundamental properties of the code-tree structure are analyzed theoretically in Section \ref{sec:prop_mode}, which are expected to be essential for constructing code trees. 
Here, we focus only on binary code symbols for simplicity. 
However, it should be noted that the proposed scheme can also be applied to the cases of non-binary code symbols.  

%%%%%%%%%%%%%%%%%%%%%%%%%%%%%%%%%%%%%%%%%%%%%%%%%%%%%%%%%%%%%%%%%%%%%%%%%%%%%%%%
\section{Preliminaries}
\label{sec:prepare}
\subsection{Notations}
The notations below are used for the following discussions. 
\begin{itemize}
	\item $\mathbb{N}$: The set of all natural numbers. 
	\item $\mathbb{R}$: The set of all real numbers. 
	\item $\mathbb{Z}^+$: The set of all non-negative integers. 
	\item $\mathbb{Z}^+_{<M}$: The set of all non-negative integers smaller than an integer $M$. 
	\item $\mathbb{A}_M$: $\{a_m\mid m\in \mathbb{Z}^+_{<M}\}$, the source alphabet of size $M$.
	\item $\mathbb{S}_M$: the Kleene closure of $\mathbb{A}_M$, or the set of all $M$-ary source symbol sequences, including a zero-length sequence $\epsilon$.
	\item $\mathbb{W}$: The set of all binary strings, including a zero-length one `$\lambda$'. `$\lambda$' can be a prefix of any binary string. 
	\item $\mathbb{M}$: $\{\textsc{Words} \subseteq \mathbb{W}\mid \textsc{Words}\neq \emptyset\}$, the set of all non-empty subsets of $\mathbb{W}$. 
	\item $\preceq$, $\npreceq$, $\prec$, $\nprec$: Dyadic relations defined in $\mathbb{W}$. $w\preceq w'\text{ (resp.~$w\npreceq w'$) }$ indicates that $w$ is (resp. is not) a prefix of $w'$. $\prec$ (resp. $\nprec$) excludes $=$ (resp. $\neq$) from $\preceq$ (resp. $\npreceq$).  
	\item $\mathbb{PF}$: $\{\textsc{Words}\in \mathbb{M}\mid \forall w\neq w'\in \textsc{Words}: w\npreceq w'\}$, the set of all prefix-free binary string sets. 
	\item $\parallel$: A dyadic relation defined for $\mathbb{W}$ or $\mathbb{M}$. For $\mathbb{W}$, $w\parallel w'$ indicates that $w$ and $w'$ satisfy either $w\preceq w'$ or $w'\preceq w$. If $w\npreceq w'$ and $w'\npreceq w$, we write $w\nparallel w'$. For $\mathbb{M}$, $\textsc{Words}\parallel \textsc{Words}'$ indicates that there are some $w\in \textsc{Words}$ and $w'\in \textsc{Words}'$ satisfying $w\parallel w'$. $\textsc{Words}\nparallel \textsc{Words}'$ means any pair of $w\in \textsc{Words}$ and $w'\in \textsc{Words}'$ is $w\nparallel w'$.
	\item $\oslash$: An operator defined for $\mathbb{W}$. $w_{\rm pre}\oslash w$ subtracts the prefix $w_{\rm pre}$ from $w$.  
	\item $\|\cdot\|_{\rm len}$: The length of a sequence in $\mathbb{S}_M$ or a string in $\mathbb{W}$.
\end{itemize}
Note that $\preceq$ is a partial order on $\mathbb{W}$, which satisfies reflexivity, antisymmetry, and transitivity, 
while $\parallel$ is a dependency relation, which satisfies reflexivity and symmetry but not transitivity. 

\subsection{Conventional binary AIFV and AIFV-$m$ codes}
\begin{figure}[!tb]
	\begin{center}
		\subfigure[Example of a binary AIFV code, equivalent to one shown in \cite{ref:aifv1}. ]{
			\includegraphics[width=6cm,  bb=0 0 537 371]{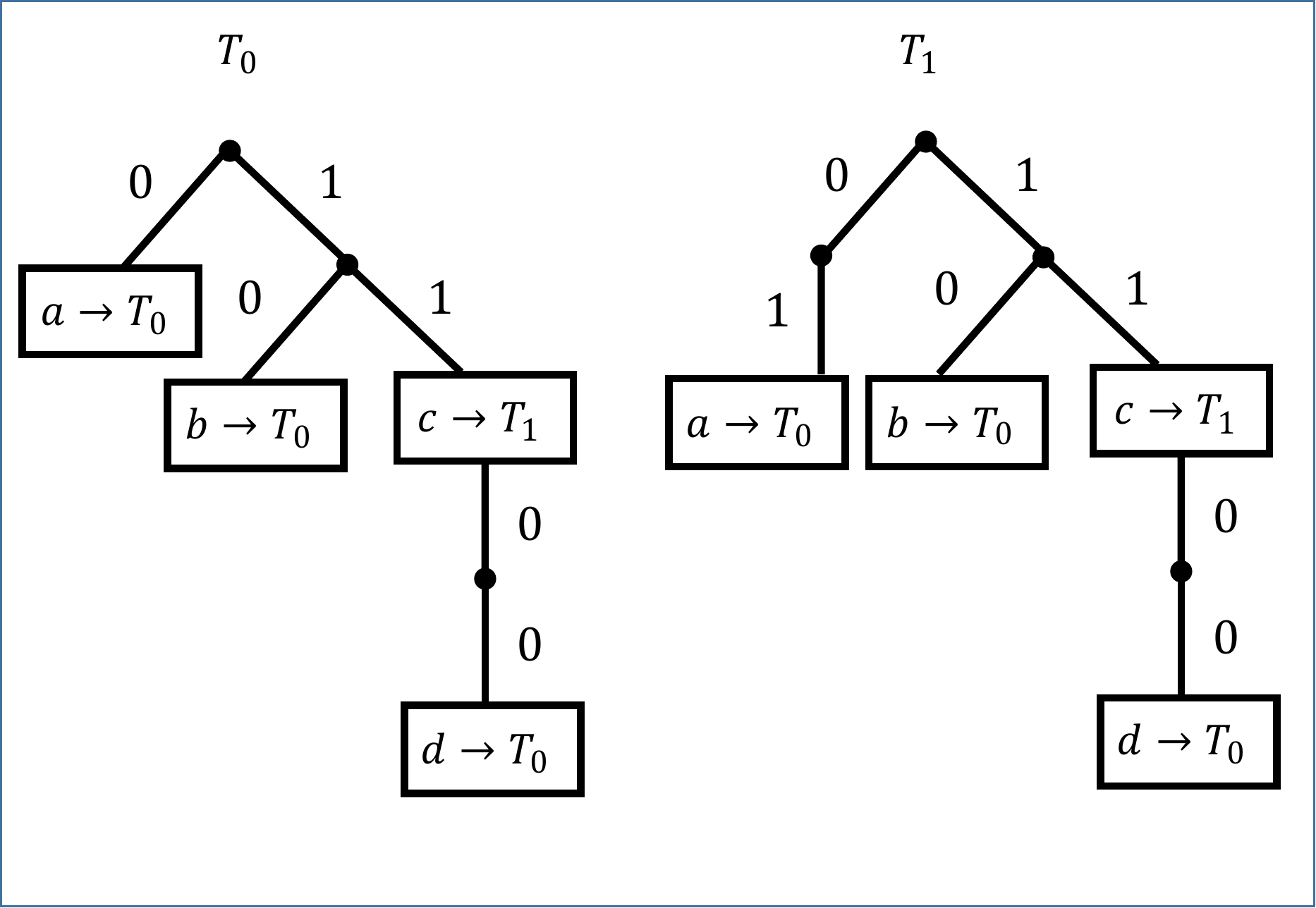}}
		\subfigure[Example of an AIFV-$m$ code, equivalent to one shown in \cite{ref:aifv2}. ]{
			\includegraphics[width=9cm,  bb=0 0 804 371]{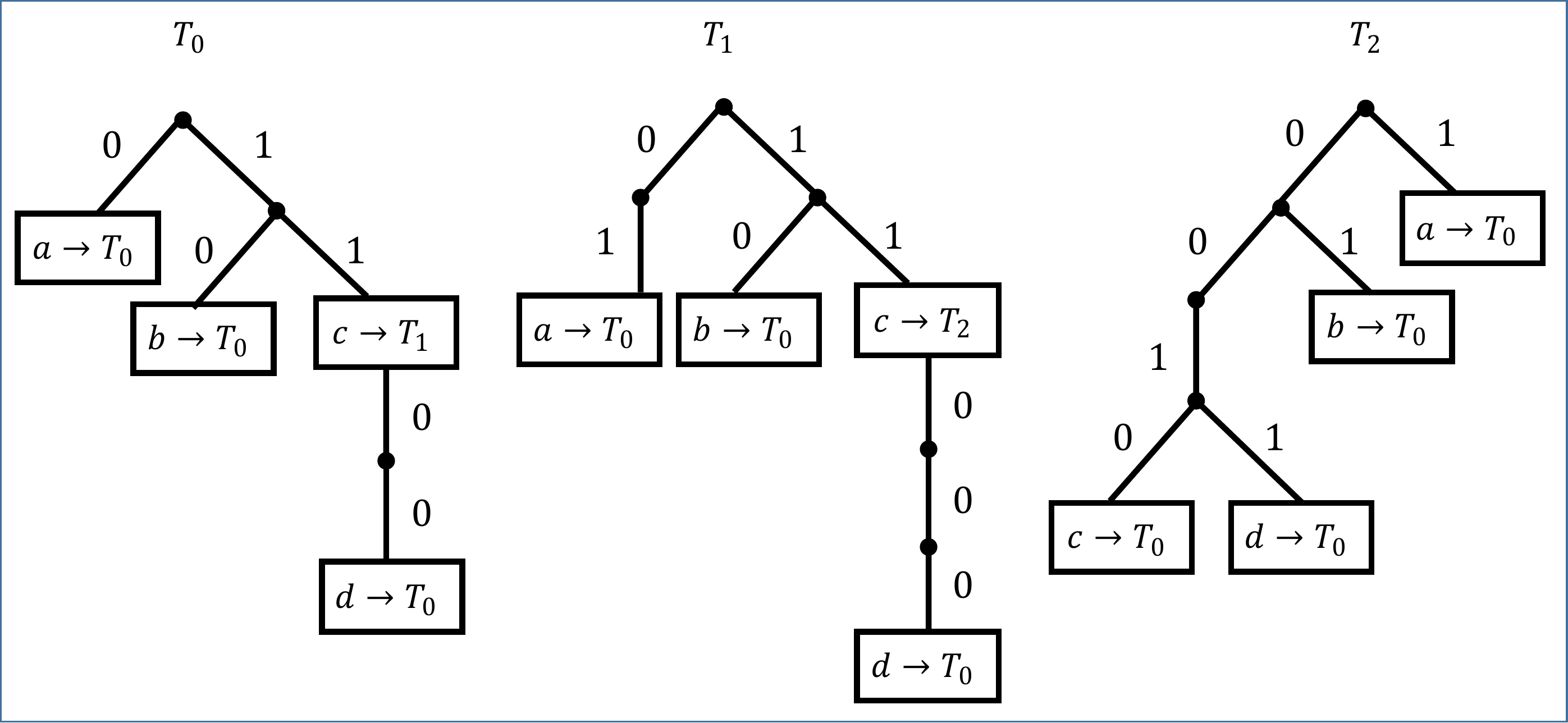}}
	\end{center}
	\caption{Examples of the conventional codes. Square boxes indicate the nodes assigned with source symbols. }
	\label{fig:aifv_ex}
\end{figure}

As we mentioned in the introduction, the conventional binary AIFV codes \cite{ref:aifv1} prepare multiple code trees to represent more flexible encoding rules. 
The code trees $T_0$ and $T_1$ are constructed to follow the rules below. 
\mytheory{rle}{Properties the code trees of binary AIFV codes must satisfy \cite{ref:aifv1}}
{
	\label{rle:aifv2}
	\begin{enumerate}
		\item Incomplete internal nodes (nodes with one child) are divided into two categories, master nodes and slave nodes. 
		\item Source symbols are assigned to either master nodes or leaves.  
		\item The child of a master node must be a slave node, and the master node is connected to its grandchild by code symbols `00'.
		\item The root of $T_1$ has two children. The child connected by `0' from the root is a slave node. The slave node is connected by `1' to its child.
	\end{enumerate} 
} 
This rule allows us to combine two code trees to a single coding rule. 
The encoding works by switching the code trees according to the previous input: 
\mytheory{prc}{Encoding a source symbol sequence into a binary AIFV codeword sequence \cite{ref:aifv1}}
{
	\label{prc:aifv2enc}	
	Follow the steps below for the input symbol sequence $x_0 x_1\cdots$. 
	\begin{enumerate}
		\item Use $T_0$ to encode the initial source symbol $x_0$.
		\item When $x_i$ is encoded by a leaf (resp. a master node), then use $T_0$ (resp. $T_1$) to encode the next symbol $x_{i+1}$. 
	\end{enumerate}	
}
Fig.~\ref{fig:aifv_ex} shows an example of a set of code trees $T_0$ and $T_1$ constructed for source symbols $\{a,b,c,d\}\in\mathbb{A}_4$. 
Under the restrictions of {\it Rule \ref{rle:aifv2}}, we can assign source symbols to internal nodes, as well as to leaves. 
In this paper, we write the code-tree switching rules in the nodes with their assigned source symbols to make them clear. 

For example, let us encode $acca$ using the code trees in Fig.~\ref{fig:aifv_ex} (a). 
The encoder starts with $T_0$ to encode $a$, outputting the codeword `0'. 
Then, it encodes $c$ by $T_0$ as `11' and switches the code tree to $T_1$. 
$T_1$ gives another `11' for $c$, with the code tree still being $T_1$. 
Similarly, `01' is output for $a$ with $T_1$ used. 
As a result, the encoded codeword sequence becomes `0111101'. 
By using two code trees, we can assign the codewords more flexibly than a single one: 
In the case of Fig.~\ref{fig:aifv_ex} (a), by introducing a slave node above the leaf assigned with $d$, a 2-bit codeword becomes available for $c$, which is impossible to assign in the binary Huffman codes for $\{a,b,c,d\}$ when $a$ is 1 bit and $b$ is 2 bits.  

The decoding also uses the code-tree switching to decode the source symbols uniquely:
\mytheory{prc}{Decoding a source symbol sequence from a binary AIFV codeword sequence \cite{ref:aifv1}}
{
	\label{prc:aifv2dec}
	Follow the steps below for the input codeword sequence. 
	\begin{enumerate}
		\item Use $T_0$ to decode the initial source symbol $x_0$.
		\item Trace the codeword sequence as long as possible from the root in the current code tree. Then, output the source symbol assigned to the reached master node or leaf.
		\item If the reached node is a leaf (resp. a master node), then use $T_0$ (resp. $T_1$) to decode the next source symbol from the current position on the codeword sequence. 
	\end{enumerate}
}
Decoding the codeword sequence in the above example goes as follows. 
The decoder starts the decoding from $T_0$, tracing the codeword sequence `0111101' as long as possible from the root. 
It reaches the leaf `0' assigned with $a$, and thus $x_0=a$. 
Then, the decoder traces the sequence `111101' according to $T_0$ again. 
Since `11' is a master node assigned with $c$ but `111' is not in $T_0$, the decoder can determine $x_1$ as $c$, switching the code tree to $T_1$. 
The next symbol is decoded from the sequence `1101'. 
`11' is a master node assigned with $c$, but `1101' is not in $T_1$. 
Therefore, $x_3$ is $c$, and $x_4$ is decoded from `01' by $T_1$. 
The decoder can reach the leaf `01' of $T_1$ assigned with $a$, and thus $x_4=a$. 
As a result, we can get the correct source symbol sequence $acca$. 

When decoding $c$ in the example, the decoder checks at most 2 bits after getting the codeword `11' for $c$ to confirm whether the encoder encoded $c$ and switched the code tree to $T_1$ or encoded $d$ instead of $c$. 
This check requires 2 bits of decoding delay for the decoder. 

As explained above, the binary AIFV code uses two code trees, permitting 2 bits of decoding delay. 
As an extension, AIFV-$m$ code is presented \cite{ref:aifv2} to tune the bit length with finer precision using more decoding delay. 
It uses $m$ code trees, permitting $m$ bits of decoding delay. 
The rules for the code trees $T_0,T_1,\cdots,T_{m-1}$ are modified from {\it Rule \ref{rle:aifv2}} as follows. 
\mytheory{rle}{Properties the code trees of AIFV-$m$ codes must satisfy \cite{ref:aifv2}}
{
	\label{rle:aifvm}
	\begin{enumerate}
		\item Any node in the code trees is either a slave node, a master node, or a complete internal node. Source symbols are only assigned to master nodes.
		\item The degree $k$ of master nodes must satisfy $0 \le k < m$. 
		\item $T_k$ ($1 \le k < m$) has a node connected to the root by a $k$-length run of zeros and is a slave-1 node.
	\end{enumerate}
} 
The terms used in this rule are defined as follows. 
\begin{itemize}
	\item Slave-0 node (resp. slave-1 node): A slave node that connects to its child by `0' (resp. `1'). 
	\item Master node of degree $k$: For $k>0$, an incomplete internal node satisfying (i) $k$ consecutive-descendant nodes are slave-0 nodes and (ii) the $(k + 1)$-th descendant is not a slave-0 node. The master nodes of degree 0 are treated as leaves.
\end{itemize}
As the binary AIFV coding, AIFV-$m$ coding introduces a code-tree switching rule to realize uniquely-decodable codes from multiple code trees: 
\mytheory{prc}{Encoding a source symbol sequence into an AIFV-$m$ codeword sequence \cite{ref:aifv2}}
{
	\label{prc:aifvmenc}	
	Follow the steps below for the input symbol sequence $x_0 x_1\cdots$. 
	\begin{enumerate}
		\item Use $T_0$ to encode the initial source symbol $x_0$.
		\item When $x_i$ is encoded by a master node of degree $k$, then use $T_k$ to encode the next symbol $x_{i+1}$.
	\end{enumerate}	
}
\mytheory{prc}{Decoding a source symbol sequence from an AIFV-$m$ codeword sequence \cite{ref:aifv2}}
{
	\label{prc:aifvmdec}
	Follow the steps below for the input codeword sequence. 
	\begin{enumerate}
		\item Use $T_0$ to decode the initial source symbol $x_0$.
		\item Trace the codeword sequence as long as possible from the root in the current code tree. Then, output the source symbol assigned to the reached master node or leaf.
		\item If the reached node is a master node of degree $k$, then use $T_k$ to decode the next source symbol from the current position on the codeword sequence.
	\end{enumerate}
}
Fig.~\ref{fig:aifv_ex} (b) gives an example of an AIFV-3 code. 
We can see that the switching rule is controlled by the number of consecutive incomplete nodes below the master node: 
$T_0$ switches to $T_1$ after encoding/decoding $c$ since there is one slave-0 node below the master node assigned with $c$; 
$T_1$ switches to $T_2$ after encoding/decoding $c$ since there are two slave-0 nodes below the master node assigned with $c$. 
When decoding $c$ by $T_1$, the decoder has to check at most 3 bits after getting the codeword `11' for $c$ to confirm whether the encoder encoded $c$ and switched the code tree to $T_2$ or encoded $d$ instead of $c$. 

Both the binary AIFV and AIFV-$m$ codes determine the switching rules by one-to-one correspondence between which code tree to switch and the number of consecutive-descendant slave-0 nodes of the nodes assigned with the source symbols: 
The code tree always switches to $T_k$ if and only if we encode/decode source symbols whose nodes have $k$ consecutive-descendant slave-0 nodes; 
the code tree always switches to $T_0$ if and only if we encode/decode the source symbols assigned to leaves. 
This restriction strongly limits the variety of codes represented by the multiple code trees. 
When we are allowed $m$ bits of decoding delay, we have to use $m$ consecutive-descendant slave-0 somewhere to fully utilize the permitted delay. 
However, it makes a large difference between the code lengths of the source symbols assigned to the master node of degree $m$ and its descendant node. 
Therefore, AIFV-$m$ codes need the source distributions to be biased if we want to enhance the compression efficiency by using larger values for $m$. 

Moreover, the conventional AIFV codes cannot represent all the codes we can uniquely decode with a given delay.
The AIFV codes presented in our previous works \cite{ref:xdgr,ref:xdg} are uniquely decodable, although they do not obey {\it Rule \ref{rle:aifvm}}. 
Note that they are designed for infinite source symbols $\mathbb{A}_{\infty}$ but can be truncated to represent AIFV codes for finite source symbols.   
In the later sections, we propose a scheme of AIFV codes to represent any code of a given decoding delay. 

\subsection{Decoding delay for general codes}
\label{sec:delay}
\subsubsection{Definition and an example for Huffman code}
\begin{figure}[t]
	\begin{center}
		\includegraphics[width=4cm,  bb=0 0 269 203]{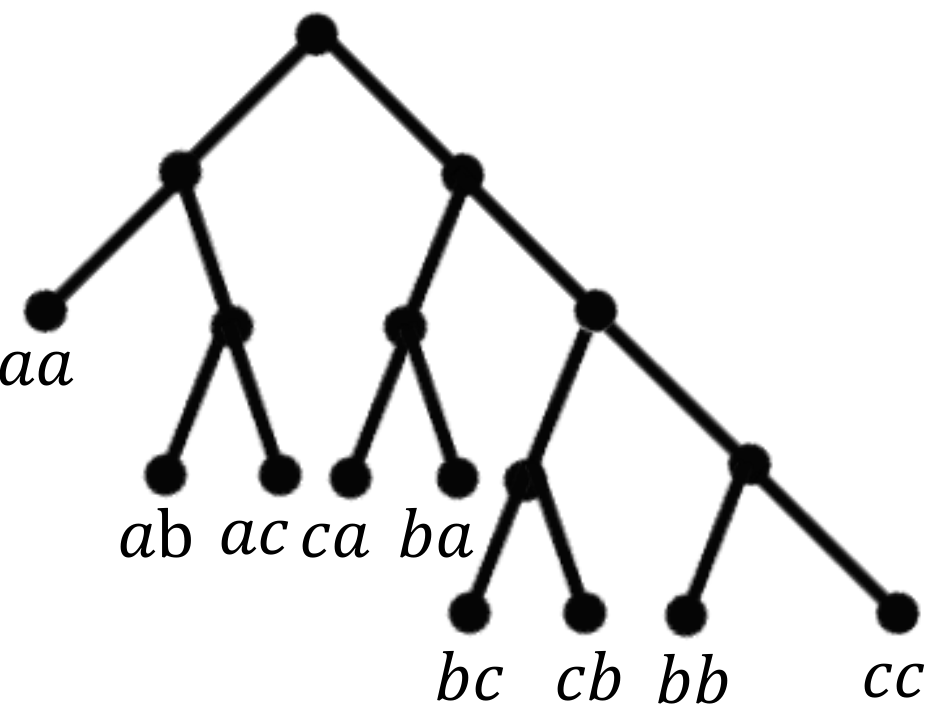}
	\end{center}
	\caption{Example of a Huffman code. }
	\label{fig:huffman_ex}
\end{figure}

For the later discussions, let us clarify the definition of delay. 
In this paper, we focus on the delay that happens in decoding each source symbol sequentially. 
\mytheory{dfn}{Decoding delay for $s\in\mathbb{S}_M$}{
	The maximum length of the binary string needed for the decoder to determine $s$ from $\{s'\in \mathbb{S}_M\mid \|s'\|_{\rm len}=\|s\|_{\rm len}\}$ as its output, after reading the codeword that the encoder can immediately determine as its output when encoding $s$. 
}
\mytheory{dfn}{Decoding delay of a code}{
	The maximum decoding delay for all $s\in\mathbb{S}_M$. 
}
This definition of delay can be applied generally. 
For example, Let us think of an extended Huffman code \cite{ref:introDC}, a Huffman code for a Cartesian product of source symbols, as in Fig.~\ref{fig:huffman_ex}. 
From now on, we omit the code symbol for each edge when two edges are connected to the same node 
and assume left-hand-side (resp.~right-hand-side) edges are for code symbol `0' (resp.~`1'). 
This code is instantaneously decodable if we interpret $\{aa, ab, ac, ba, bb, bc, ca, cb, cc\}$ as $\mathbb{A}_9$, and thus has no decoding delay for $aa, ab, ac, ba, bb, bc, ca, cb$, and $cc$. 
However, if we interpret it as a code for $\mathbb{A}_3=\{a,b,c\}$, we can break down the code to decode the source symbols sequentially and define the decoding delay for $a$, $b$, and $c$. 

In the case of Fig.~\ref{fig:huffman_ex}, the encoder can immediately determine `0' as its output when encoding $a$ because `0' is a prefix of every codeword corresponding to the sequence beginning with $a$. 
The decoder can also determine $a$ as its output when decoding `0' because every codeword having a prefix `0' corresponds to the symbol sequences beginning with $a$.
Therefore, the decoding delay for the source symbol $a$ is zero. 
On the other hand, the encoder can immediately determine `1' as its output when encoding $b$ because `1' is a prefix of every codeword corresponding to the sequence beginning with $b$. 
However, the decoder cannot determine $b$ as its output only by getting `1' because the codewords corresponding to the sequences beginning with $c$ also begin with `1'. 
For this sake, the decoder has to read `01', `100', or `110' after `1' to determine $b$ as its output. 
Similarly, the decoder has to read `00', `101', or `111' to determine $c$ as its output. 
Therefore, the extended Huffman code of this example is decodable with a decoding delay of 3 bits. 

\subsubsection{Leading and following codewords}
When thinking of a code with a non-symbol-wise coding rule like the extended Huffman code, its decoding delay depends on how we break down a code to make symbol-wise decoding, and in general, the way of breaking down a code is not unique: 
The example stated above for Fig.~\ref{fig:huffman_ex} shows one way of breaking down the extended Huffman code, 
but it is not the only way to get a symbol-wise coding rule. 
Although it is more awkward, we can say, for instance, that the encoder can immediately determine `$\lambda$' as its output when encoding $b$ because `$\lambda$' is a prefix of every codeword corresponding to the sequence beginning with $b$ in Fig.~\ref{fig:huffman_ex}. 
In this case, the decoder has to read `101', `1100', or `1110' after `$\lambda$' to determine $b$ as its output. 
When breaking down the code in this way and implementing symbol-wise encoder and decoder according to it, the decoding delay becomes 4 bits. 
The important fact for the main discussion is that we can break down a code in some way, even if it is not defined in a symbol-wise manner. 

Considering the above fact, we generalize the idea to arbitrary VV codes. 
Say $V:\mathbb{S}_M\to\mathbb{W}$ is some VV code. 
We define some terms using $V(s)$ for any source symbol sequence $s(\in\mathbb{S}_M)$: 
\begin{itemize}
	\item $\textrm{Lcword}_V(s)$: Leading codeword, a codeword that the encoder of $V$ can immediately determine as its output when encoding $s$, regardless of its succeeding source symbols. In other words, one picked up from $\{w\in \mathbb{W}\mid \forall \textrm{Tail}\in\mathbb{S}_M: w\preceq V(s\textrm{Tail})\}$. It is arbitrarily selected for each $s$. By definition, $\textrm{Lcword}_V(\epsilon)=$`$\lambda$'. 
	\item $\textrm{Fcword}_V(s| \textrm{Tail})$: Following codeword, a codeword that follows the leading codeword $\textrm{Lcword}_V(s)$ when encoding $s\textrm{Tail}$ by $V$. In other words, one picked up from $\{w\mid \textrm{Lcword}_V(s)w\preceq V(s\textrm{Tail}), w\in \mathbb{W}\}$. It is  arbitrarily selected for each $(s, \textrm{Tail})$. 
\end{itemize} 
When a code $V(s)$ is given, we have some choices of which binary string to set as $\textrm{Lcword}_V(s)$ for each $s$. 
The leading codeword $\textrm{Lcword}_V(s)$ can be any binary string that forms a common prefix of all encoded words staring with $s$. 
For given $V(s)$ and $\textrm{Lcword}_V(s)$, we also have some choices for $\textrm{Fcword}_V(s| \textrm{Tail})$. 
The following codeword $\textrm{Fcword}_V(s| \textrm{Tail})$ can be any binary string that makes $\textrm{Lcword}_V(s)\textrm{Fcword}_V(s| \textrm{Tail})$ a prefix of $V(s\textrm{Tail})$ and enables the decoder to decode $s$ from $\textrm{Lcword}_V(s)\textrm{Fcword}_V(s| \textrm{Tail})$. 

Using the notation above, we can write as follows the condition where the decoder of $V$ can determine $s$ as its output. 
\begin{equation}
	\label{eq:lead_follow_cond}
	\forall s'\neq s, \|s'\|_{\rm len}=\|s\|_{\rm len}, \forall \textrm{Tail}, \textrm{Tail}'\in\mathbb{S}_M: \textrm{Lcword}_V(s)\textrm{Fcword}_V(s|\textrm{Tail}) \nparallel \textrm{Lcword}_V(s')\textrm{Fcword}_V(s'|\textrm{Tail}').
\end{equation}
If $V(s)$ is decodable with a decoding delay of $N$ bits, it means we can set such leading and following codewords with the length of the following codeword not longer than $N$ bits. 
For uniquely decodable $V(s)$, we can always set some leading and following codewords satisfying Eq.~(\ref{eq:lead_follow_cond}) because, in the worst case, setting $\textrm{Fcword}_V(s|\textrm{Tail})=\textrm{Lcword}_V(s)\oslash V(s\textrm{Tail})$ will do. 
Note that $\textrm{Lcword}_V(s)$ can take `$\lambda$' so that we can always set the leading codewords as well. 

\begin{table}[tb]
	\caption{Example of the leading and following codewords for the code in Fig.~\ref{fig:huffman_ex}. }
	\begin{center}
		\begin{tabular}{|c|c|c|c|c|c|c|c|c|}
			\cline{1-4} \cline{6-9}
			$s_1$ &  $\textrm{Lcword}_V(s_1)$   & $s_2$ & $\textrm{Fcword}_V(s_1|s_2)$ &  & $s_1$ &  $\textrm{Lcword}_V(s_1)$   & $s_2$ & $\textrm{Fcword}_V(s_1|s_2)$          \\ \cline{1-4} \cline{6-9} 
			\multirow{3}{*}{$a$} & \multirow{3}{*}{`0'} & $a$ & \multirow{3}{*}{`$\lambda$'} &  & $aa$ & `00' & \multirow{9}{*}{$\epsilon$} & \multirow{9}{*}{`$\lambda$'} \\ \cline{3-3} \cline{6-7}
			&                   & $b$ &  &  & $ab$ & `010' &                   &                   \\ \cline{3-3} \cline{6-7}
			&                   & $c$ &  &  & $ac$ & `011' &                   &                   \\ \cline{1-4} \cline{6-7}
			\multirow{3}{*}{$b$} & \multirow{3}{*}{`1'} & $a$ & `01' &  & $ba$ & `101' &                   &                   \\ \cline{3-4} \cline{6-7}
			&                   & $b$ & `110' &  & $bb$ & `1110' &                   &                   \\ \cline{3-4} \cline{6-7}
			&                   & $c$ & `100' &  & $bc$ & `1100' &                   &                   \\ \cline{1-4} \cline{6-7}
			\multirow{3}{*}{$c$} & \multirow{3}{*}{`1'} & $a$ & `00' &  & $ca$ & `100' &                   &                   \\ \cline{3-4} \cline{6-7}
			&                   & $b$ & `101' &  & $cb$ & `1101' &                   &                   \\ \cline{3-4} \cline{6-7}
			&                   & $c$ & `111' &  & $cc$ & `1111' &                   &                   \\ \cline{1-4} \cline{6-9} 
		\end{tabular}
	\end{center}
	\label{tb:huffman_lfcw}
\end{table}

For the code of Fig.~\ref{fig:huffman_ex}, we can set the leading and following codewords as in Table \ref{tb:huffman_lfcw}.
Owing to this setting, for example, we can retrieve $a$ from $\textrm{Lcword}_V(a)\textrm{Fcword}_V(a|a)=\textrm{Lcword}_V(a)\textrm{Fcword}_V(a|b)=\textrm{Lcword}_V(a)\textrm{Fcword}_V(a|c)=$`0' and $b$ from $\textrm{Lcword}_V(b)\textrm{Fcword}_V(b|a)$=`101', $\textrm{Lcword}_V(b)\textrm{Fcword}_V(b|b)$=`1110', or $\textrm{Lcword}_V(b)\textrm{Fcword}_V(b|c)$=`1100'. 

Moreover, by breaking down the codewords, we can determine the codewords for an arbitrary length sequence. 
In the case of Fig.~\ref{fig:huffman_ex}, the codewords are defined for each source symbol pair, and thus usually, we cannot encode sequences of odd lengths. 
However, the decoder can retrieve the source symbol sequentially if we have properly-set leading and following codewords. 
So, if the odd-length sequence ends with $b$, we can encode it into the leading codeword `1' and one of the following codewords, say `01'. 
For instance, $V(aaa)$ for the code in Fig.~\ref{fig:huffman_ex} can be given by `000', `00' from $V(aa)$ and `0' from $\textrm{Lcword}_V(a)\textrm{Fcword}_V(a|a)$. 
$V(caabc)$ can be written as `100010100', `100' from $V(ca)$, `010' from $V(ab)$, and `100' from $\textrm{Lcword}_V(c)\textrm{Fcword}_V(c|a)$. 
Of course, the decoder has to know the exact length of the source symbol sequence if it wants to retrieve exactly $caabc$ rather than $caabca$. 
However, at least the decoder never mistakes the retrieval of the $caabc$ part. 
In this case, the codeword $\textrm{Fcword}_V(c|a)$ is what we call the termination codeword in the later discussion. 

It should be noted that, in this paper, we do not focus on how to find the leading and following codeword for a given code. 
Our purpose is only to show that the decoding delay is a general idea for VV codes.

\subsubsection{Example for VF code}
\begin{figure}[t]
	\begin{center}
		\includegraphics[width=4cm,  bb=0 0 285 185]{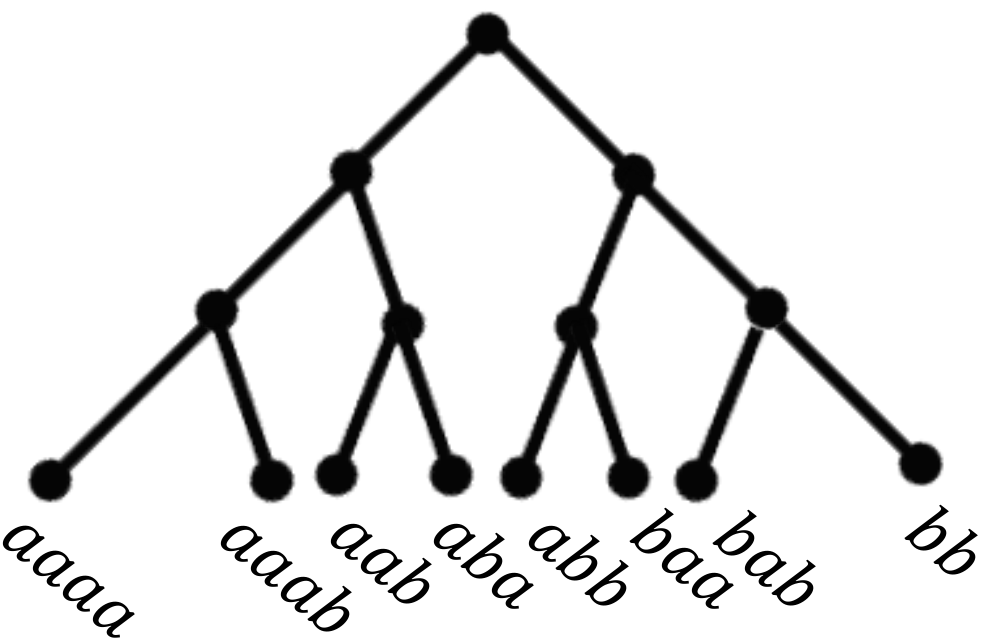}
	\end{center}
	\caption{Example of a VF code. }
	\label{fig:tunstall_ex}
\end{figure}

\begin{table}[tb]
	\caption{Example of the leading and following codewords for the code in Fig.~\ref{fig:tunstall_ex}. }
	\begin{center}
		\begin{tabular}{|c|c|c|c|c|c|c|c|c|}
			\cline{1-4} \cline{6-9}
			$s_1$                & $\textrm{Lcword}_V(s_1)$ & $s_2$ & $\textrm{Fcword}_V(s_1|s_2)$ &  & $s_1$                 & $\textrm{Lcword}_V(s_1)$ & $s_2$      & $\textrm{Fcword}_V(s_1|s_2)$ \\ \cline{1-4} \cline{6-9} 
			\multirow{5}{*}{$a$} & \multirow{5}{*}{`$\lambda$'} & $aaa$ & \multirow{4}{*}{`0'}         &  & \multirow{3}{*}{$aa$} & \multirow{3}{*}{`0'}     & $aa$       & \multirow{2}{*}{`0'}         \\ \cline{3-3} \cline{8-8}
			&                          & $aab$ &                              &  &                       &                          & $ab$       &                              \\ \cline{3-3} \cline{8-9} 
			&                          & $ab$  &                              &  &                       &                          & $b$        & `10'                         \\ \cline{3-3} \cline{6-9} 
			&                          & $ba$  &                              &  & \multirow{2}{*}{$ab$} & \multirow{2}{*}{`$\lambda$'} & $a$        & `011'                        \\ \cline{3-4} \cline{8-9} 
			&                          & $bb$  & `100'                        &  &                       &                          & $b$        & `100'                        \\ \cline{1-4} \cline{6-9} 
			\multirow{3}{*}{$b$} & \multirow{3}{*}{`1'}     & $aa$  & `01'                         &  & \multirow{2}{*}{$ba$} & \multirow{2}{*}{`1'}     & $a$        & `01'                         \\ \cline{3-4} \cline{8-9} 
			&                          & $ab$  & \multirow{2}{*}{`1'}         &  &                       &                          & $b$        & `10'                         \\ \cline{3-3} \cline{6-9} 
			&                          & $b$   &                              &  & $bb$                  & `111'                    & $\epsilon$ & `$\lambda$'                      \\ \cline{1-4} \cline{6-9} 
		\end{tabular}
	\end{center}
	\label{tb:tunstall_lfcw}
\end{table}

Some may think that the above discussion supports only FV codes. 
However, it can also be applied to Variable-to-Fixed-length (VF) codes. 
Fig.~\ref{fig:tunstall_ex} shows an example. 
In this case, the codewords corresponding to the sequences beginning with $a$ are `000', `001', `010', `011', and `100'. 
Since there are codewords starting with `0' and with `1', the encoder cannot determine any codeword by only getting $a$, so the leading codeword for $a$ is $\textrm{Lcword}_V(a)=$`$\lambda$'. 
However, every codeword starting with `0' corresponds to $a\cdots$, and so the decoder can retrieve $a$ from $0$. 
Of course, it can also decode $a$ from `100'. 
Therefore, we can set `0' and `100' as the following codewords. 

Similarly, we can set the leading and following codewords as in Table \ref{tb:tunstall_lfcw}. 
Using these codewords, we can define the codewords for any length of source sequence: 
For instance, $V(baba)=$`1100', `110' from $V(bab)$ and `0' from $\textrm{Lcword}_V(a)\textrm{Fcword}_V(a|aaa)$; 
$V(ababa)=$`011101', `011' from $V(aba)$ and `101' from $\textrm{Lcword}_V(ba)\textrm{Fcword}_V(ba|a)$. 

%%%%%%%%%%%%%%%%%%%%%%%%%%%%%%%%%%%%%%%%%%%%%%%%%%%%%%%%%%%%%%%%%%%%%%%%%%%%%%%%
\section{Proposed $N$-bit-delay AIFV codes}
\label{sec:prop}
\subsection{Basic structure}
Based on the decoding delay defined in the previous section, we propose a scheme of AIFV codes that can represent all codes we can decode within a given amount of delay. 
To show the proposed scheme, we first introduce a term ``mode,'' the core concept of the proposed codes. 
Modes play a critical role in representing the rules of the code tree structure. 
As explained later, the mode defined here is used as sets of allowed prefixes for a code tree and works as a query for the decoder to determine which tree is used. 
However, we want to clarify the whole class before discussing the consistency of the code. 
Therefore, we define the mode simply as follows. 
\mytheory{dfn}{Mode of a code tree}
{
	An arbitrary member of $\mathbb{M}$ assigned to a code tree. 
}
We redefine the code trees by assigning a mode $\textsc{Mode}$ ($\in \mathbb{M}$) to each one:
\begin{itemize}
	\item $\mathbb{T}_M = \{(\textrm{Cword}_k, \textrm{Point}_k, \textsc{Mode}_k, k)\mid \textrm{Cword}_k: \mathbb{A}_{M}\to \mathbb{W}, \textrm{Point}_k: \mathbb{A}_{M}\to \mathbb{Z}^+, \textsc{Mode}_k\in \mathbb{M}, k\in \mathbb{Z}^+\}$, a set of all code trees for source symbols $\mathbb{A}_M$. 
\end{itemize} 
Here, $\textrm{Cword}_k(a)$ and $\textrm{Point}_k(a)$ are respectively the codeword and the index of the next code tree corresponding to the source symbol $a$. 
$k$ is the index of the code tree, and throughout this paper, we indicate it by a subscript like in $T_k$. 

A code-tree set $\{T_k\}$ is represented here as an element of a subset of $\mathbb{T}_M$ whose code trees are all available for encoding and decoding processes:
\begin{itemize}
	\item $\mathbb{RT}_M = \{\{T_k\mid k\in \mathbb{Z}^+_{<K}\}\subset\mathbb{T}_M\mid K\in\mathbb{N}$, $T_k$ is reachable from $T_0$ for $k\neq0$, $T_k$ may not reach $T_{k'}$ of $k'\notin\mathbb{Z}^+_{<K}\}$, a set of all reachable (and closed) code-tree sets. 
\end{itemize} 
The word ``reachable'' is used here similarly to the context of Markov chains \cite{ref:markovchain}: 
$T_k$ is reachable from $T_0$, or $T_0$ may reach $T_k$, when the encoder can switch the code tree to $T_k$ from $T_0$ within finite steps following the switching rules.  
Note that $\mathbb{T}_M$ and $\mathbb{RT}_M$ only clarify the components of code trees and code-tree sets without discussing their decodability. 

\begin{figure}[tb]
	\begin{center}
		\includegraphics[width=13cm,  bb=0 0 799 186]{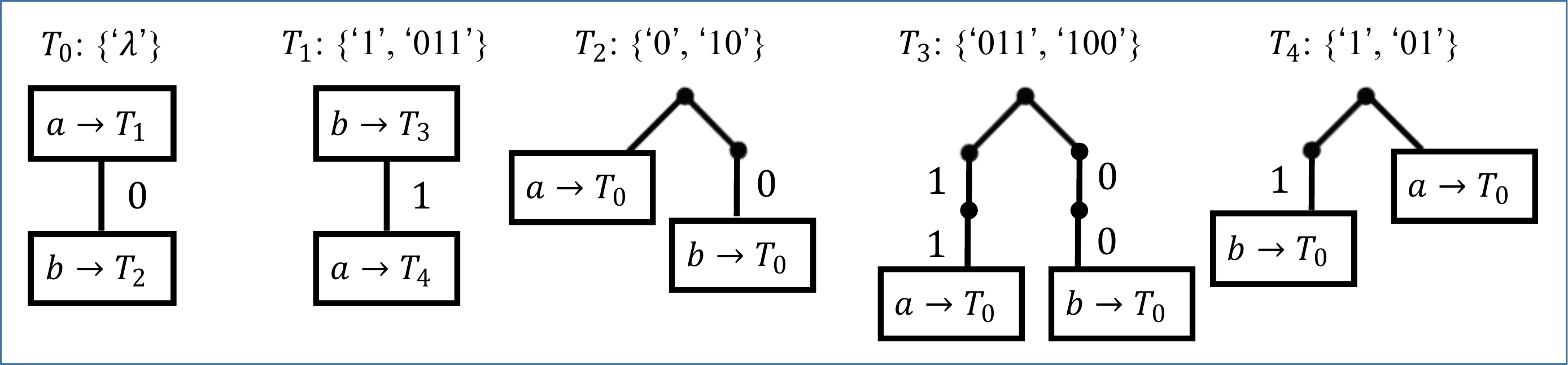}
	\end{center}
	\caption{An example of a code-tree set $\{T_k\}$ of the proposed code, with modes \{`$\lambda$'\}, \{`1', `011'\}, \{`0', `10'\}, \{`011', `100'\}, and \{`1', `01'\} for the respective code trees. }
	\label{fig:trees_ex}
\end{figure}

As shown by an example in Fig.~\ref{fig:trees_ex}, each code tree determines, for every source symbol, the codeword and the next code tree. 
The code-tree set can be equivalently written in the form of a table as in Table \ref{tb:codebook} and is used for the encoding and decoding procedures as follows. 
\mytheory{prc}{Encoding a source symbol sequence into a proposed AIFV codeword sequence}
{
	\label{prc:enc}	
	Follow the steps below with the $L$-length source symbol sequence $x_0 x_1\cdots x_{L-1}$ ($\in \mathbb{S}_M$) and code-tree set $\{T_k\}$ ($\in \mathbb{RT}_M$) being the inputs of the encoder. 
	\begin{enumerate}
		\item Start encoding from $k=0$.
		\item For $i=0,1,\cdots, L-1$, output the codeword $\textrm{Cword}_k(x_i)$ in the current code tree $T_k$
		and switch the code tree by updating the index $k$ with $\textrm{Point}_k(x_i)$.
		\item Output some binary string in the mode $\textsc{Mode}_k$ (here, we call it the termination codeword). 
	\end{enumerate}	
}
As proven later, the termination codeword can be an arbitrary member of $\textsc{Mode}_k$. 
However, from a practical perspective, we use the one having minimum length unless otherwise specified. 
\mytheory{prc}{Decoding a source symbol sequence from a proposed AIFV codeword sequence}
{
	\label{prc:dec}
	Follow the steps below with the codeword sequence, code-tree set $\{T_k\}$, and output length $L$ being the inputs of the decoder. 
	\begin{enumerate}
		\item Start decoding from $k=0$. 
		\item Compare the codeword sequence with the codewords in the current code tree $T_k$. 
		If the codeword $\textrm{Cword}_k(a)$ matches the codeword sequence, 
		and some codeword $\textrm{Query}\in \textsc{Mode}_{\textrm{Point}_k(a)}$ matches the codeword sequence after $\textrm{Cword}_k(a)$, 
		output the source symbol $a$ and continue the process from the codeword sequence right after $\textrm{Cword}_k(a)$. 
		\item Switch the code tree by updating the index $k$ with $\textrm{Point}_k(a)$. 
		\item If the decoder has output less than $L$ symbols, return to b. 
	\end{enumerate}
}
For example, think of encoding a source symbol sequence $abbaa$ using the code trees in Fig.~\ref{fig:trees_ex}. 
The encoder starts with $T_0$ to encode $a$, outputting the codeword `$\lambda$'. 
Then, it switches the code tree to $T_1$ to encode $b$, outputting another `$\lambda$' and switching the code tree to $T_3$. 
$T_3$ gives the codeword `100' for $b$ and switches the code tree to $T_0$. 
Similarly, it respectively outputs codewords `$\lambda$' and `1' using $T_0$ and $T_1$ for $a$ and another $a$. 
For the code's termination, the encoder outputs the minimum-length binary string in the mode of $T_4$, `1'. 
As a result, the encoded codeword sequence becomes `$\lambda\lambda$100$\lambda$11', namely `10011'. 

The decoder starts the decoding from $T_0$, checking at first whether the codeword `$\lambda$' of $a$ matches the codeword sequence `10011'. 
The codeword for $a$ is `$\lambda$' and thus matches the sequence. 
Then, the decoder checks whether any codeword in the mode of $T_1$, the code tree $a$ points, matches the sequence. 
Since `1' is included, it outputs $a$ and switches the code tree to $T_1$. 
The next symbol is decoded from the codeword sequence following `$\lambda$', i.e., `10011'. 
The codeword `$\lambda$' for $b$ in $T_1$ matches the sequence, and the following `100' is included in the mode of $T_3$. 
Therefore, $b$ is output, and the third symbol is decoded from `10011' by $T_3$. 
The codeword `100' is for $b$ in $T_3$, and `$\lambda$' in the mode of $T_0$ obviously matches the following sequence so that another $b$ is output, with the fourth symbol decoded from `11' by $T_0$. 
Similarly, the decoder outputs $a$ and decodes the fifth symbol from `11' by $T_1$. 
Although `11' matches the codeword `$\lambda$' of the symbol $b$ in $T_1$, no codeword in the mode of $T_3$ matches `11'. 
Thus, the decoder does not output $b$ and instead checks the codeword for $a$. 
Since the codeword `1' of $a$ matches the sequence and the following `1' is included in the mode of $T_4$, the decoder can determine the last source symbol $a$ to output. 
As a result, we can get the correct source symbol sequence $abbaa$. 

The termination codewords in step c of the encoding are necessary when the decoder only knows the total length $L$ of the source symbol sequence and cannot know the end of the codeword sequence. 
We can easily understand their role by thinking of encoding a single $a$ using the code trees in Fig.~\ref{fig:trees_ex}. 
The code tree $T_0$ gives `$\lambda$', and thus if there is no termination codeword and the decoder does not know the end of the codeword sequence, it starts to check the following irrelevant binary strings: 
When some binary string unrelated to the AIFV codeword sequence begins with `000' and follows the encoded `$\lambda$', even if the decoder knows it has to decode only one source symbol, it checks the following `000' and outputs $b$. 
If the encoder outputs the termination codeword `1' after `$\lambda$', the decoder checks it, outputs $a$ correctly, and stops the decoding process before reading the following irrelevant binary strings. 

The code defined by Fig.~\ref{fig:trees_ex}, gives the codewords `11', `101', `011', `011', `100', `00', and `010' respectively for the source symbols sequences $aaa\cdots$, $aab\cdots$, $aba\cdots$, $abb\cdots$, $ba\cdots$, and $bb\cdots$. 
It is effective for sources where $a$ appears a little more frequently than $b$, which the conventional AIFV codes cannot effectively compress because it requires the difference of the code lengths of $a$ and $b$ to be 3 bits in some code tree to utilize the allowed 3-bit decoding delay. 

Indeed, the table size increases by $|\{T_k\}|$ times compared to the binary Huffman codes. 
However, the proposed AIFV codes can assign codewords more flexibly to source symbol sequences with simple encoding/decoding processes: 
The difference between Huffman coding in the encoding process is only the symbol-wise switching of the code trees; 
the decoding process requires only at most $M(=|\mathbb{A}_M|)$ times of additional check of codewords in the modes. 
Of course, the computational complexity depends on how to implement the processes. 
For example, if we implement the decoding process as a finite-state automaton, it needs only one check for each source symbol. 

\begin{table}[t]
	\tiny
	\caption{Table for the code in Fig.~\ref{fig:trees_ex}. }
	\begin{center}
		\begin{tabular}{c|cc|cc|cc|cc|cc|}
			\cline{2-11}
			& \multicolumn{2}{c|}{\shortstack{\\$T_0$: \{`$\lambda$'\}}}    & \multicolumn{2}{c|}{$T_1$: \{`1', `011'\}}    & \multicolumn{2}{c|}{$T_2$: \{`0', `10'\}}    & \multicolumn{2}{c|}{$T_3$: \{`011', `100'\}}    & \multicolumn{2}{c|}{$T_4$: \{`1', `01'\}}    \\ \hline
			\multicolumn{1}{|c|}{\shortstack{\\Source}} & \multicolumn{1}{l|}{Codeword} & Next tree & \multicolumn{1}{l|}{Codeword} & Next tree & \multicolumn{1}{l|}{Codeword} & Next tree & \multicolumn{1}{l|}{Codeword} & Next tree & \multicolumn{1}{l|}{Codeword} & Next tree \\ \hline
			\multicolumn{1}{|c|}{$a$} & \multicolumn{1}{c|}{`$\lambda$'} & \shortstack{\\$T_1$} & \multicolumn{1}{c|}{`1'} & $T_4$ & \multicolumn{1}{c|}{`0'} & $T_0$ & \multicolumn{1}{c|}{`011'} & $T_0$ & \multicolumn{1}{c|}{`1'} & $T_0$ \\ \hline
			\multicolumn{1}{|c|}{$b$} & \multicolumn{1}{c|}{`0'} & \shortstack{\\$T_2$} & \multicolumn{1}{c|}{`$\lambda$'} & $T_3$ & \multicolumn{1}{c|}{`10'} & $T_0$ & \multicolumn{1}{c|}{`100'} & $T_0$ & \multicolumn{1}{c|}{`01'} & $T_0$ \\ \hline
		\end{tabular}
	\end{center}
	\label{tb:codebook}
\end{table}

\begin{figure}[t]
	\begin{center}
		\includegraphics[width=5cm,  bb=0 0 293 148]{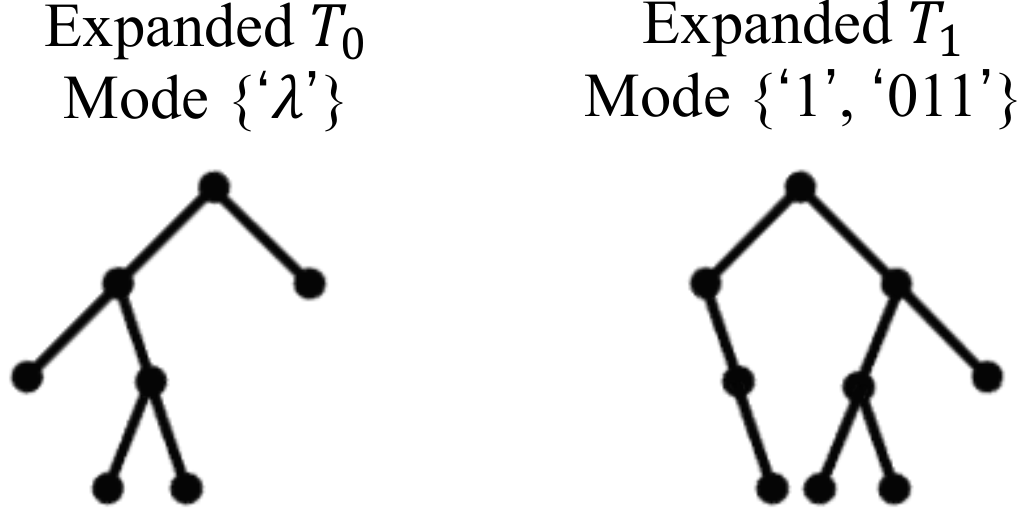}
	\end{center}
	\caption{Expanded trees representing the expanded codewords of $T_0$ and $T_1$ in Fig.~\ref{fig:trees_ex}. }
	\label{fig:expand_ex}
\end{figure}

As is evident from the example, the proposed AIFV codes have no one-to-one correspondence between the switching rules and the code tree structure: 
For example, in $T_0$, the code tree switches to $T_2$ instead of $T_0$ when encoding $b$ even though it is assigned to the leaf. 
This fact allows for much more flexible code design than the conventional ones. 
The binary strings in the modes work as queries suggesting which code tree the encoder switched. 
To discuss the rule for the code tree construction, let us define the idea of expanding codewords. 
\mytheory{dfn}{Expanded codewords for $a$ of a code tree $T_k$}
{
	$\textsc{Expand}_k(a)=\{\textrm{Cword}_k(a)\textrm{Query}\mid \textrm{Query}\in\textsc{Mode}_{\textrm{Point}_k(a)}\}$
}
\mytheory{dfn}{Expanded codeword set of a code tree $T_k$}
{
	$\textsc{Expands}_k=\{\textsc{Expand}_k(a)\mid a\in\mathbb{A}_M\}$
}
For example, in Table \ref{tb:codebook}, $T_0$ has codewords $\textrm{Cword}_0(a)=$ `$\lambda$' and $\textrm{Cword}_0(b)=$ `0' with the corresponding modes $\textsc{Mode}_{\textrm{Point}_0(a)}=\{$`1', `011'$\}$ and $\textsc{Mode}_{\textrm{Point}_0(b)}=\{$`0', `10'$\}$. 
In this case, the sets of expanded codewords of $T_0$ for $a$ and $b$ are $\textsc{Expand}_0(a)=\{$`1', `011'$\}$ and $\textsc{Expand}_0(b)=\{$`00', `010'$\}$, respectively. 

The rule to make a code-tree set $\{T_k\}$ representing a uniquely-decodable code is written as
\mytheory{rle}{Constraints for the proposed AIFV codes to be uniquely decodable}
{
	\label{rle:decodable}
	\begin{enumerate}
		\item $\forall k, \forall a\neq a': \textsc{Expand}_k(a)\nparallel \textsc{Expand}_k(a')$. 
		\label{srle:prefix_free_expansion}
		\item $\forall k, \forall \textrm{Expcw} \in \textsc{Expands}_k$, $\exists \textrm{Query} \in \textsc{Mode}_k$: $\textrm{Query} \preceq \textrm{Expcw}$. 
		\label{srle:prefix_in_mode}
	\end{enumerate} 
}
{\it Rule \ref{rle:decodable}} \ref{srle:prefix_free_expansion} means that, for any code tree, the sets of expanded codewords are prefix-free to each other. 
{\it Rule \ref{rle:decodable}} \ref{srle:prefix_in_mode} requires, for any code tree, that every expanded codeword has a prefix being a member of its mode. 
We can represent the constraints in a code-tree-wise way as above by using the idea of expanded codewords, constructed by the binary strings in the modes of the code trees pointed. 
Fig.~\ref{fig:expand_ex} shows an example of expanded trees that represent all the expanded codewords of $T_0$ and $T_1$ in Fig.~\ref{fig:trees_ex}. 
It is clear that both of them are prefix-free. 
The mode of $T_0$ is $\{$`$\lambda$'$\}$, and thus every expanded codeword of $T_0$ has a prefix in its mode. 
The expanded codewords of $T_1$ are `011', `100', `101', and `11', having `1' or `011' ($\in\textsc{Mode}_1$) as a prefix. 

We use the following notation for the discussions below to represent the constraints. 
\begin{itemize}
	\item $\mathbb{DT}_M = \{\{T_k\}\in \mathbb{RT}_M\mid \{T_k\}$ satisfies {\it Rule \ref{rle:decodable}} $\}$, a set of all reachable code-tree sets meeting {\it Rule \ref{rle:decodable}}. 
\end{itemize}

\subsection{Decodability}
Unique decodability is guaranteed for the encoding and decoding procedures stated above: 
\mytheory{thm}{Decodability of the proposed AIFV codes}
{
	\label{thm:decodable}
	For any code-tree set $\{T_k\}\in \mathbb{DT}_M$, the codeword sequence given by {\it Procedure \ref{prc:enc}} using $\{T_k\}$ is uniquely decodable by {\it Procedure \ref{prc:dec}} using $\{T_k\}$.  
} 
\underline{{\it Proof}}: 
Suppose the encoding algorithm in {\it Procedure \ref{prc:enc}} encodes an $L$-length source symbol sequence $x_0x_1\cdots x_{L-1}(\in \mathbb{S}_M)$ using code trees $T_{k_0},T_{k_1},\cdots,T_{k_{L-1}}$, respectively. 
The encoded codeword sequence becomes $w_{\rm total}=\textrm{Cword}_{k_0}(x_0)\textrm{Cword}_{k_1}(x_1)\cdots \textrm{Cword}_{k_{L-1}}(x_{L-1})\textrm{Query}_{k_L}$. 
Here, $\textrm{Query}_{k_L}$ is the termination codeword, a member of $\textsc{Mode}_{k_L}$ where $k_L=\textrm{Point}_{k_{L-1}}(x_{L-1})$. 
We claim {\it Procedure \ref{prc:dec}} can retrieve $x_0x_1\cdots x_{L-1}$ from $w_{\rm total}$ when $\{T_k\}$ satisfies {\it Rule \ref{rle:decodable}}. 
We will prove it inductively, showing the decoding algorithm in the $i$-th iteration uses the code tree $T_{k_i}$ and decodes $x_i$ correctly.  

i) [Base case] {\it Procedure \ref{prc:dec}} starts at $T_0=T_{k_0}$. 
Since $k_1=\textrm{Point}_{k_0}(x_0)$ and since $\{T_k\}$ satisfies {\it Rule \ref{rle:decodable}} \ref{srle:prefix_in_mode}, there is a binary string $\textrm{Query}_{k_1}\in \textsc{Mode}_{k_1}$ being a prefix of $\textrm{Cword}_{k_1}(x_1)\cdots \textrm{Cword}_{k_{L-1}}(x_{L-1})\textrm{Query}_{k_L}$. 
Thus, $\textrm{Cword}_{k_0}(x_0)\textrm{Query}_{k_1}\preceq w_{\rm total}$. 
Because of {\it Rule \ref{rle:decodable}} \ref{srle:prefix_free_expansion}, the expanded codeword $\textrm{Cword}_{k_0}(x_0)\textrm{Query}_{k_1}(\in \textsc{Expand}_{k_0}(x_0))$ is prefix-free among the expanded codewords in $\textsc{Expand}_{k_0}(a)$ for any other symbol $a\neq x_0$ of $T_{k_0}$. 
So the decoder can retrieve $x_0$ uniquely from $\textrm{Cword}_{k_0}(x_0)\textrm{Query}_{k_1}$, switching the code tree to $T_{k_1}$. 

ii) [Induction step] Suppose the decoder uses $T_{k_i}$ in the $i$-th iteration. 
Since $k_{i+1}=\textrm{Point}_{k_i}(x_i)$ and since $\{T_k\}$ satisfies {\it Rule \ref{rle:decodable}} \ref{srle:prefix_in_mode}, there is a binary string $\textrm{Query}_{k_{i+1}}\in \textsc{Mode}_{k_{i+1}}$ which is a prefix of $\textrm{Cword}_{k_{i+1}}(x_{i+1})\cdots \textrm{Cword}_{k_{L-1}}(x_{L-1})\textrm{Query}_{k_L}$. 
Thus, $\textrm{Cword}_{k_{i}}(x_{i})\textrm{Query}_{k_{i+1}}\preceq \textrm{Cword}_{k_{i}}(x_{i})\textrm{Cword}_{k_{i+1}}(x_{i+1})\cdots \textrm{Cword}_{k_{L-1}}(x_{L-1})\textrm{Query}_{k_L}$. 
Because of {\it Rule \ref{rle:decodable}} \ref{srle:prefix_free_expansion}, the expanded codeword $\textrm{Cword}_{k_{i}}(x_{i})\textrm{Query}_{k_{i+1}}(\in \textsc{Expand}_{k_{i}}(x_i))$ is prefix-free among the expanded codewords in $\textsc{Expand}_{k_{i}}(a)$ for any other symbol $a\neq x_i$ of $T_{k_i}$, and so the decoder can retrieve $x_i$ uniquely from $\textrm{Cword}_{k_i}(x_i)\textrm{Query}_{k_{i+1}}$, switching the code tree to $T_{k_{i+1}}$. $\qquad \blacksquare$

As we mentioned before, we can use any termination codeword as long as it is a member of $\textsc{Mode}_k$. 
In fact, the minimum-length one should be used to make the encoded codeword sequence as short as possible. 

\subsection{Decoding delay}
We can determine the decoding delay of the proposed AIFV codes by checking their modes: 
\mytheory{thm}{Decoding delay of the proposed AIFV codes}{
	Decoding delay of the code given by {\it Procedure \ref{prc:enc}} and {\it Procedure \ref{prc:dec}} with $\{T_k\}\in \mathbb{DT}_M$ is  
	\begin{equation}
		\max_k \left(\max \{\|\textrm{Query}_k\|_{\rm len}\mid\exists \textrm{Expcw} \in \textsc{Expands}_k: \textrm{Query}_k\preceq \textrm{Expcw}, \textrm{Query}_k\in \textsc{Mode}_k\}\right).
		\label{eq:delay}
	\end{equation}
}
\underline{{\it Proof}}: As in the previous proof, suppose the encoding algorithm in {\it Procedure \ref{prc:enc}} encodes an $L$-length source symbol sequence $x_0x_1\cdots x_{L-1}(\in \mathbb{S}_M)$ using code trees $T_{k_0},T_{k_1},\cdots,T_{k_{L-1}}$, respectively. 
During {\it Procedure \ref{prc:enc}}, the leading codeword when encoding $x_0x_1\cdots x_{i-1}$ ($i\in \mathbb{Z}^+_{<L}$) corresponds to $\textrm{Cword}_{k_0}(x_0)\textrm{Cword}_{k_1}(x_1)\cdots \textrm{Cword}_{k_{i-1}}(x_{i-1})$. 
The following codeword needed for the decoder to determine $x_0x_1\cdots x_{i-1}$ as its output is $\textrm{Query}_{k_{i}}$, which is a member of $\textsc{Mode}_{k_{i}}$. 
Since $\textrm{Query}_{k_{i}}\preceq \textrm{Cword}_{k_{i}}(x_{i})\textrm{Query}_{k_{i+1}} (\in \textsc{Expands}_{k_{i}})$ from {\it Rule \ref{rle:decodable}} \ref{srle:prefix_in_mode}, even if some binary string is in $\textsc{Mode}_{k_{i}}$, 
it would not be output in {\it Procedure \ref{prc:enc}} as $\textrm{Query}_{k_{i}}$ unless it is a prefix of some expanded codeword of $T_{k_{i}}$. 
Therefore, the decoding delay of the code is given by the maximum length of the binary string in any mode being the prefix of some expanded codeword. 
$\qquad \blacksquare$\\\\

The theorem reveals that the decoding delay heavily depends on the modes of code trees. 
Note that when determining the decoding delay of the proposed AIFV codes, we have to check whether the members of the modes are actually used as the prefixes of the expanded codewords, as well as to check their lengths. 
From now on, we define the proposed AIFV codes as follows. 
\mytheory{dfn}{$N$-bit-delay AIFV code}{
	The code given by {\it Procedure \ref{prc:enc}} using a code-tree set $\{T_k\}\in \mathbb{DT}_M$ with modes having codewords of at most $N$-bit-length.
}
It should be noted that $0$-bit-delay AIFV codes are identical to instantaneous FV codes, including Huffman codes: They can be interpreted as codes using code-tree sets of size 1 and $\{$`$\lambda$'$\}$ for the mode. 
The relationship between the conventional AIFV-$m$ and the proposed codes is explained in the later section. 

\subsection{Generality}
We claim that any VV code, decodable within a finite delay, can be constructed as a set of code trees in the proposed scheme. 
\mytheory{thm}{Generality of the proposed $N$-bit-delay AIFV codes}{
	\label{thm:generality}
	For any uniquely encodable and uniquely decodable VV code $V:\mathbb{S}_M\to\mathbb{W}$ which can be decoded with a decoding delay of $N$ bits, there is an $N$-bit-delay AIFV code giving a codeword $V(s)$ for any $s\in\mathbb{S}_M$. 
}
\underline{{\it Proof}}: We show here that we can rewrite $V(s)$ equivalently into a code-tree set satisfying the constraints of $N$-bit-delay AIFV codes. 
From the assumption, $V(s)$ is decodable with a decoding delay of $N$ bits, and thus we can define the leading and following codewords $\textrm{Lcword}_V(s)$ and $\textrm{Fcword}_V(s|\textrm{Tail})$ for any $s, \textrm{Tail}\in\mathbb{S}_M$ satisfying Eq.~(\ref{eq:lead_follow_cond}) and $\|\textrm{Fcword}_V(s|\textrm{Tail})\|_{\rm len} \leq N$. 

Without loss of generality, we can set $\textrm{Lcword}_V(s)$ and $\textrm{Fcword}_V(sa|\textrm{Tail})$ to satisfy the following. 
\begin{equation}
	\label{eq:lead_follow_for_whole_string}
	\textrm{Lcword}_V(s)\textrm{Fcword}_V(s|\epsilon)=V(s),
\end{equation}
\begin{equation}
	\label{eq:lead_prefix}
	\forall a\in\mathbb{A}_M: \textrm{Lcword}_V(s)\preceq \textrm{Lcword}_V(sa),
\end{equation}
\begin{equation}
	\label{eq:exp_prefix}
	\forall a\in\mathbb{A}_M: \textrm{Lcword}_V(s)\textrm{Fcword}_V(s|a\textrm{Tail})\preceq \textrm{Lcword}_V(sa)\textrm{Fcword}_V(sa|\textrm{Tail}).
\end{equation}

[Reason for Eq.~(\ref{eq:lead_follow_for_whole_string})] Setting $\textrm{Fcword}_V(s|\epsilon)=\textrm{Lcword}_V(s)\oslash V(s)$ does not conflict with the definition of the following codeword $\textrm{Fcword}_V$. 

[Reason for Eqs.~(\ref{eq:lead_prefix}) and (\ref{eq:exp_prefix})] Let us think by dividing the conditions. 
We can say from the definitions of $\textrm{Lcword}_V$ and $\textrm{Fcword}_V$ that $\textrm{Lcword}_V(s)\parallel \textrm{Lcword}_V(sa)$ and $\textrm{Lcword}_V(s)\textrm{Fcword}_V(s|a\textrm{Tail})\parallel \textrm{Lcword}_V(sa)\textrm{Fcword}_V(sa|\textrm{Tail})$. 

i) If $\textrm{Lcword}_V(sa)\prec \textrm{Lcword}_V(s)$ and $\textrm{Lcword}_V(s)\textrm{Fcword}_V(s|a\textrm{Tail})\preceq \textrm{Lcword}_V(sa)\textrm{Fcword}_V(sa|\textrm{Tail})$, we can reset $\textrm{Lcword}_V(sa)$ and $\textrm{Fcword}_V(sa|\textrm{Tail})$ as 
\begin{eqnarray}
	\label{eq:update_lfcw_1_1}
	\textrm{Lcword}_V(sa)&\equiv& \textrm{Lcword}_V(s) \\
	\label{eq:update_lfcw_1_2}
	\textrm{Fcword}_V(sa|\textrm{Tail})&\equiv& \textrm{Lcword}_V(s)\oslash(\textrm{Lcword}_V(sa)\textrm{Fcword}_V(s|a\textrm{Tail})). 
\end{eqnarray}
Eq.~(\ref{eq:update_lfcw_1_1}) obeys the definition of $\textrm{Lcword}_V(sa)$. 
This operation does not change the codeword of $\textrm{Lcword}_V(sa)\textrm{Fcword}_V(sa|\textrm{Tail})$, and thus Eq.~(\ref{eq:lead_follow_cond}) still holds. 
Even if $\textrm{Tail}=\epsilon$, $\textrm{Lcword}_V(sa)\textrm{Fcword}_V(sa|\epsilon)=V(sa)$ still holds, too. 
Additionally, Eq.~(\ref{eq:update_lfcw_1_2}) shortens $\textrm{Fcword}_V(sa|\textrm{Tail})$ so that $\|\textrm{Fcword}_V(sa|\textrm{Tail})\|_{\rm len} < N$. 

ii) If $\textrm{Lcword}_V(s)\preceq \textrm{Lcword}_V(sa)$ and $\textrm{Lcword}_V(sa)\textrm{Fcword}_V(sa|\textrm{Tail})\prec \textrm{Lcword}_V(s)\textrm{Fcword}_V(s|a\textrm{Tail})$, we can reset $\textrm{Fcword}_V(s|a\textrm{Tail})$ as 
\begin{equation}
	\label{eq:update_lfcw_2}
	\textrm{Fcword}_V(sa|\textrm{Tail})\equiv \textrm{Lcword}_V(s)\oslash(\textrm{Lcword}_V(sa)\textrm{Fcword}_V(s|a\textrm{Tail})). 
\end{equation}
This operation shortens $\textrm{Fcword}_V(sa|\textrm{Tail})$ and gives $\textrm{Lcword}_V(s)\textrm{Fcword}_V(s|a\textrm{Tail})=\textrm{Lcword}_V(sa)\textrm{Fcword}_V(sa|\textrm{Tail})$. 
Since we can retrieve $sa$ from $\textrm{Lcword}_V(sa)\textrm{Fcword}_V(sa|\textrm{Tail})$, we can of course retrieve $s$ from $\textrm{Lcword}_V(s)\textrm{Fcword}_V(s|a\textrm{Tail})$, and so Eq.~(\ref{eq:lead_follow_cond}) still holds. 

iii) If $\textrm{Lcword}_V(sa)\prec \textrm{Lcword}_V(s)$ and $\textrm{Lcword}_V(sa)\textrm{Fcword}_V(sa|\textrm{Tail})\prec \textrm{Lcword}_V(s)\textrm{Fcword}_V(s|a\textrm{Tail})$, we can reset $\textrm{Lcword}_V(s)$ and $\textrm{Fcword}_V(s|a\textrm{Tail})$ as 
\begin{eqnarray}
	\label{eq:update_lfcw_3_1}
	\textrm{Lcword}_V(s)&\equiv& \textrm{Lcword}_V(sa) \\
	\label{eq:update_lfcw_3_2}
	\textrm{Fcword}_V(s|a\textrm{Tail})&\equiv& \textrm{Fcword}_V(sa|\textrm{Tail}). 
\end{eqnarray}
Eq.~(\ref{eq:update_lfcw_3_1}) only shortens $\textrm{Lcword}_V(s)$ so that it does not disturb the definition of $\textrm{Lcword}_V(s)$. 
Eq.~(\ref{eq:update_lfcw_3_2}) does not make the length of $\textrm{Fcword}_V(s|a\textrm{Tail})$ longer than $N$. 
Similarly to ii), Eq.~(\ref{eq:lead_follow_cond}) still holds. 
Therefore, the conditions in Eqs.~(\ref{eq:lead_prefix}) and (\ref{eq:exp_prefix}) do not disturb the generality. \\

Using the above fact, we can equivalently represent the VV code by setting the code trees as follows. 
\begin{equation}
	\label{eq:vvcodetree}
	T_{k_s}=(\textrm{Cword}_{k_s}, \textrm{Point}_{k_s}, \textsc{Mode}_{k_s}, k_s),
\end{equation}
where
\begin{eqnarray}
	\textrm{Cword}_{k_s}(a)&=&\textrm{Lcword}_V(s)\oslash\textrm{Lcword}_V(sa),\\
	\textrm{Point}_{k_s}(a)&=&k_{sa},\\
	\textsc{Mode}_{k_s}&=&\{\textrm{Fcword}_V(s|\textrm{Tail})\mid \textrm{Tail}\in \mathbb{S}_M\}\equiv\textsc{Follow}_V(s).
\end{eqnarray}
$k_s$ is a non-negative integer defined for $s\in \mathbb{S}_M$ and being $k_s\neq k_{s'}$ for $s\neq s'$. 
Especially, $k_\epsilon = 0$. 

If we use $\textrm{Fcword}_V(s|\epsilon)$ ($\in \textsc{Follow}_V(s)$) when encoding $s$, the codeword sequence for $s$ given by $\{T_k\}$ is $\textrm{Lcword}_V(s)\textrm{Fcword}_V(s|\epsilon)$, which is $V(s)$ according to Eq.~(\ref{eq:lead_follow_for_whole_string}). 
Therefore, $\{T_k\}$ gives equivalent codewords as $V$. 

The expanded codewords of $T_{k_s}$ are 
\begin{equation}
	\textsc{Expand}_{k_s}(a)=\{\textrm{Lcword}_V(s)\oslash\textrm{Lcword}_V(sa)\textrm{Fcword}_V(sa|\textrm{Tail})\mid \textrm{Tail}\in \mathbb{S}_M\}.
\end{equation}
From Eq.~(\ref{eq:lead_follow_cond}), 
\begin{equation}
	\forall a'\neq a, \forall \textrm{Tail}, \textrm{Tail}'\in\mathbb{S}_M: \textrm{Lcword}_V(sa)\textrm{Fcword}_V(sa|\textrm{Tail}) \nparallel \textrm{Lcword}_V(sa')\textrm{Fcword}_V(sa'|\textrm{Tail}'), 
\end{equation}
which becomes, by using Eq.~(\ref{eq:lead_prefix}), 
\begin{eqnarray}
	\forall a'\neq a, \forall \textrm{Tail}, \textrm{Tail}'\in\mathbb{S}_M:&&\nonumber\\ \textrm{Lcword}_V(s)\oslash\textrm{Lcword}_V(sa)\textrm{Fcword}_V(sa|\textrm{Tail}) &\nparallel& \textrm{Lcword}_V(s)\oslash\textrm{Lcword}_V(sa')\textrm{Fcword}_V(sa'|\textrm{Tail}')\nonumber\\
	\iff\forall a'\neq a, \textsc{Expand}_{k_s}(a)&\nparallel& \textsc{Expand}_{k_s}(a'). 
\end{eqnarray}
Thus, $\{T_k\}$ satisfies {\it Rule \ref{rle:decodable}} \ref{srle:prefix_free_expansion}. 
On the other hand, from Eq.~(\ref{eq:exp_prefix}), we have 
\begin{equation}
	\forall a\in\mathbb{A}_M: \textrm{Fcword}_V(s|a\textrm{Tail})\preceq \textrm{Lcword}_V(s)\oslash\textrm{Lcword}_V(sa)\textrm{Fcword}_V(sa|\textrm{Tail}),
\end{equation}
where $\textrm{Fcword}_V(s|a\textrm{Tail})\in \textsc{Follow}_V(s)$ and $\textrm{Lcword}_V(s)\oslash\textrm{Lcword}_V(sa)\textrm{Fcword}_V(sa|\textrm{Tail})\in\textsc{Expand}_{k_s}(a)$
Therefore, {\it Rule \ref{rle:decodable}} \ref{srle:prefix_in_mode} holds. 
Additionally, every codeword in the mode $\textsc{Follow}_V(s)$ is used as a prefix of the expanded codeword in $\textsc{Expand}_{k_s}(a)$. 
According to the assumption of decoding delay of the VV code, every codeword in the mode $\textsc{Follow}_V(s)$ is not longer than $N$ bits. 
So the code trees in Eq.~(\ref{eq:vvcodetree}) construct an $N$-bit-delay AIFV code. 
$\qquad \blacksquare$\\\\

This theorem reveals that there exists an $N$-bit-delay AIFV code for any VV code if we set an appropriate decoding delay $N$. 
The proof above also implies that we can make the modes from the following codewords. 
Therefore, the modes can be inferred from code-tree sets as ones of the conventional AIFV codes, whose modes are not defined. 

\section{Properties of code-tree modes}
\label{sec:prop_mode}
\subsection{Basic modes}
\begin{figure}[t]
	\begin{center}
		\includegraphics[width=12cm,  bb=0 0 748 228]{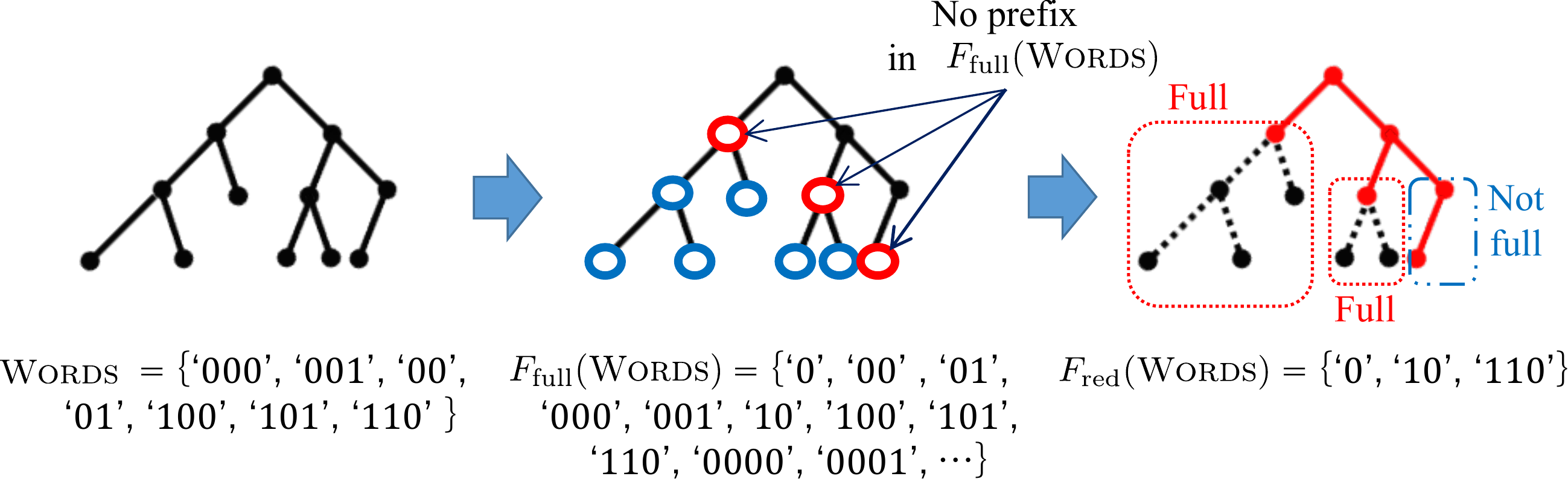}
	\end{center}
	\caption{Example of a reduced binary string set. A tree representing $\textsc{Words}$ on the left, circled nodes illustrating the members of $F_{\rm full}(\textsc{Words})$ in the middle, and the tree of $F_{\rm red}(\textsc{Words})$ on the right.}
	\label{fig:reduce_ex}
\end{figure}

According to the definition of the mode, any binary string set can be a mode of a code tree as long as {\it Rule \ref{rle:decodable}} is satisfied. 
However, many code trees are meaningless: We can replace the codewords and modes to shorten the decoding delay for any code trees without changing the encoding output. 
To discuss the basic pattern of the modes, we introduce the following notation.
\begin{itemize}
	\item $\sim$: A dyadic relation defined for $\mathbb{DT}_M$. $\{T_k\}\sim \{T'_k\}$ indicates that for every source symbol sequence, the codeword sequences given by respectively using $\{T_k\}$ and $\{T'_k\}$ become identical when their termination codewords are truncated appropriately. 
\end{itemize}
This relation groups code-tree sets neglecting the trivial difference due to termination codewords.  
Say we encode $ab$ using $\{T_k\}$ as `0' for $a$ by $T_0$, `$\lambda$' for $b$ by $T_1$, and a termination codeword `100' in $T_2$, which results in `0100'. 
If we have $\{T'_k\}$ that gives `0' for $a$ by $T'_0$, `1' for $b$ by $T'_1$, and a termination codeword `0' in $T'_2$, which results in `010', 
the codewords for $ab$ can be identical by truncating the end of the termination codeword `100' in $T_2$. 
If we can always make the outputs identical by truncating the end of the termination codeword, we write as $\{T_k\}\sim \{T'_k\}$. 

The following functions are also introduced. 
\begin{itemize}
	\item $f_{\rm cmn}$: $\mathbb{M}\to \mathbb{W}$. $f_{\rm cmn}(\textsc{Words})$ outputs the maximum-length common prefix of $\textsc{Words} \in \mathbb{M}$. 
	\item $F_{\rm full}$: $\mathbb{M}\to\mathbb{M}$. $F_{\rm full}(\textsc{Words})=\{\textrm{Prefix}\in\mathbb{W}\mid \forall \textrm{Suffix}\in\mathbb{W}, \exists w\in \textsc{Words}: \textrm{Prefix}\:\textrm{Suffix}\parallel w\}$.
	\item $F_{\rm red}$: $\mathbb{M}\to\mathbb{PF}$. $F_{\rm red}(\textsc{Words})=\{\hat{w}\in F_{\rm full}(\textsc{Words})\mid \forall \textrm{Prefix}\in F_{\rm full}(\textsc{Words}): \textrm{Prefix}\nprec \hat{w}\}$.  
\end{itemize}
It is easier to understand $F_{\rm full}(\textsc{Words})$ and $F_{\rm red}(\textsc{Words})$ as operations on trees. 
For a tree given by $\textsc{Words}\in\mathbb{M}$, $F_{\rm full}(\textsc{Words})$ finds the nodes that have full trees below, and $F_{\rm red}(\textsc{Words})$ cuts off such trees. 

Fig.~\ref{fig:reduce_ex} provides an example. 
$F_{\rm full}(\textsc{Words})$ checks for every string whether we can make it prefix-free from $\textsc{Words}$ by adding some suffix: 
If we make some string $\hat{w}=$ `0$\:\textrm{Suffix}$' by any $\textrm{Suffix}\in\mathbb{W}$, there is always a string $w\in \textsc{Words}$ that is $\hat{w}\parallel w$, and thus `0' $\in F_{\rm full}(\textsc{Words})$; for `1', we can make `111' $\nparallel w(\in \textsc{Words})$, and thus `1' $\notin F_{\rm full}(\textsc{Words})$. 
As a result, $F_{\rm full}(\textsc{Words})$ contains the codewords corresponding to the circled nodes shown in the middle of Fig.~\ref{fig:reduce_ex}. 
$F_{\rm full}(\textsc{Words})$ also contains the descendants of the leaves in $\textsc{Words}$. 
Then, $F_{\rm red}(\textsc{Words})$ picks up only the nodes which have no prefix in $F_{\rm full}(\textsc{Words})$, namely $\{$`0', `10', `110'$\}$. 
Eventually, every full subtree in $\textsc{Words}$ gets reduced in the tree represented by $F_{\rm red}(\textsc{Words})$. 

Note that $F_{\rm red}(\textsc{Words})$ is defined for $\textsc{Words}\in \mathbb{M}$, which includes binary string sets not satisfying the prefix condition. 
When we interpret $F_{\rm red}(\textsc{Words})$ as an operation of reducing trees, we do not have to consider strings in $\textsc{Words}$ which have a prefix included in $\textsc{Words}$. 
For example, in Fig.~\ref{fig:reduce_ex}, if we have also `$11$' as a member of $\textsc{Words}$, the tree below `$\lambda$' becomes a full tree. 
This is because `$11$' added to $\textsc{Words}$ is a prefix of `$\lambda 111$', and we cannot make any prefix-free binary string starting with `$\lambda$'. 
Due to the definition of $F_{\rm red}(\textsc{Words})$, the reduced binary string set is always prefix-free, even if $\textsc{Words}$ is not. 

\begin{figure}[!tb]
	\begin{center}
		\subfigure[Example of a code-tree set $\{T_k\}\in\mathbb{DT}_3$ with modes $\{$`$\lambda$'$\}$, $\{$`000', `001', `010'$\}$, and $\{$`011', `100', `101'$\}$. ]{
			\includegraphics[width=8.5cm,  bb=0 0 703 233]{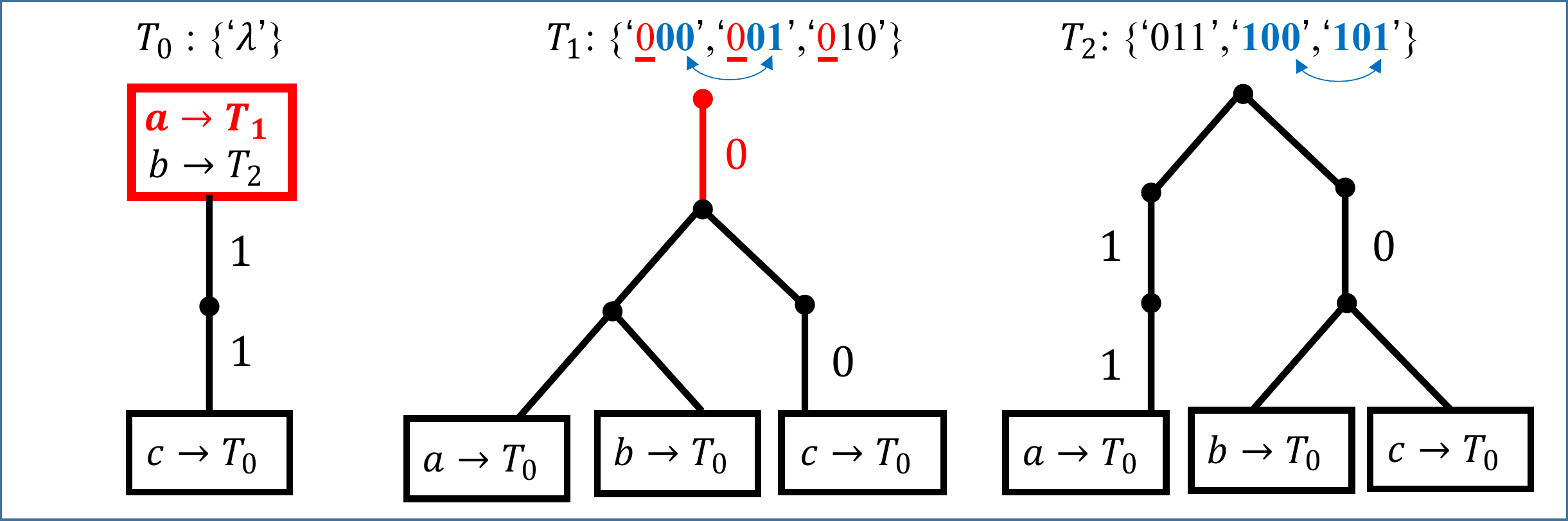}}
		\subfigure[The code-tree set $\{\tilde{T}_k\}$ with the basic modes $\{$`$\lambda$'$\}$, $\{$`0', `10'$\}$, and $\{$`011', `10'$\}$. ]{
			\includegraphics[width=8.5cm,  bb=0 0 703 233]{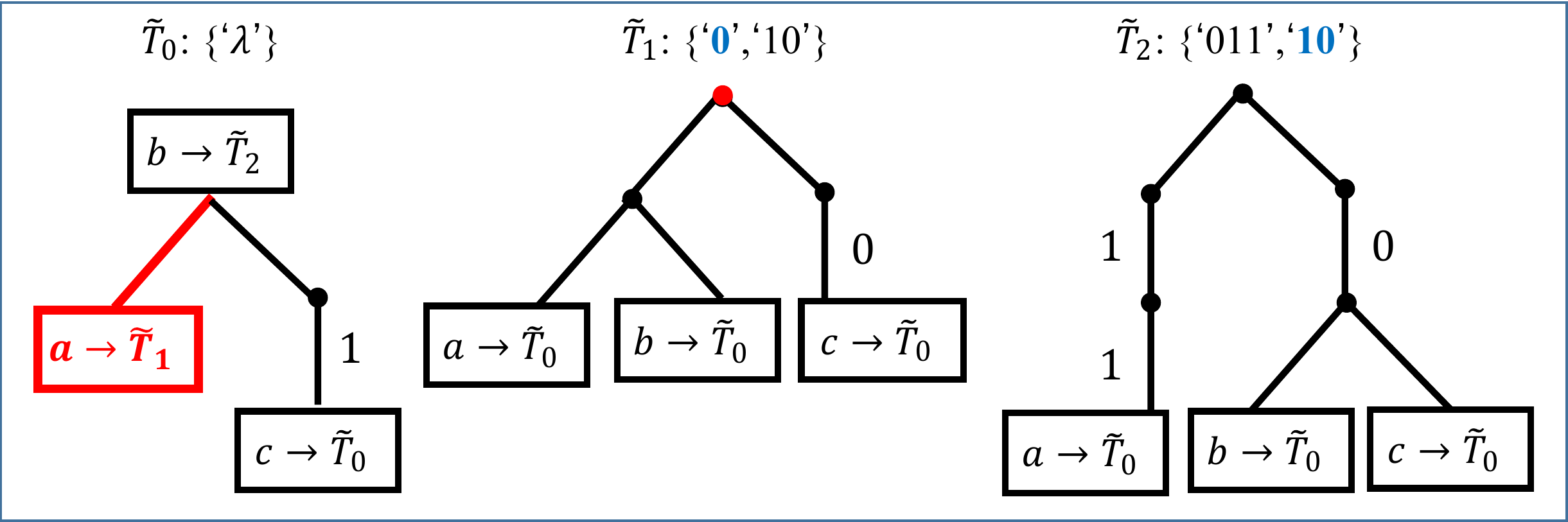}}
	\end{center}
	\caption{Example of constructing a code-tree set $\{\tilde{T}_k\}\sim \{T_k\}$ using only the basic modes. }
	\label{fig:reduced_ex}
\end{figure}

The functions $f_{\rm cmn}(\textsc{Words})$ and $F_{\rm red}(\textsc{Words})$ help us represent some essential features of the code-tree sets and modes:
\mytheory{dfn}{Full code-tree set}{
	A code-tree set where each code tree $T_k$ has a mode $\textsc{Mode}_k$ and an expanded codeword set $\textsc{Expand}_k(a)$ satisfying $F_{\rm red}(\textsc{Mode}_k)=F_{\rm red}(\textsc{Expands}_k)$ and especially $\textsc{Mode}_0=\{$`$\lambda$'$\}$. 
}
\mytheory{dfn}{Basic mode}{
	A mode $\{f_{\rm cmn}(\textsc{Words})\oslash\hat{\textrm{Query}}\mid \hat{\textrm{Query}}\in F_{\rm red}(\textsc{Words})\}$ using arbitrary $\textsc{Words}\in \mathbb{M}$. 
}
Full code-tree sets are contained with code trees whose expanded codewords cannot satisfy {\it Rule \ref{rle:decodable}} if any single string is added to them. 
Basic mode is a class of modes having no common prefix except `$\lambda$' and being invariable by $F_{\rm red}$, which play an important role in defining representative modes for code trees. 
Let us define a conversion using the basic modes: 
\mytheory{prc}{Conversing a code-tree set into one with basic modes only}
{
	\label{prc:basic}	
	For a given code-tree set $\{T_k = (\textrm{Cword}_k, \textrm{Point}_k, \textsc{Mode}_k, k)\}$, output 
	\begin{equation}
		\label{eq:Ttilde}
		\{\tilde{T}_k = (\tilde{\textrm{Cword}}_k, \textrm{Point}_k, \tilde{\textsc{Mode}}_k, k)\},
	\end{equation}	
	where
	\begin{eqnarray}
		\label{eq:conv1}
		\tilde{\textsc{Mode}}_k &=& \{f_{\rm cmn}(\textsc{Mode}_k)\oslash\hat{\textrm{Query}}_k\mid \hat{\textrm{Query}}_k\in F_{\rm red}(\textsc{Mode}_k)\}\\
		\label{eq:conv2}
		\tilde{\textrm{Cword}}_k(a) &=& f_{\rm cmn}(\textsc{Mode}_k)\oslash \textrm{Cword}_k(a)f_{\rm cmn}(\textsc{Mode}_{\textrm{Point}_k(a)}).
	\end{eqnarray}
}
The code-tree sets made by {\it Procedure \ref{prc:basic}} show the following properties. 
\mytheory{thm}{Equivalence}{
	\label{thm:equiv}
	$\{\tilde{T}_k\}$ given by $\{T_k\}\in \mathbb{DT}_M$ using {\it Procedure \ref{prc:basic}} satisfies $\{\tilde{T}_k\}\sim \{T_k\}$ when $\textsc{Mode}_0=\{$`$\lambda$'$\}$. 
}
\mytheory{thm}{Minimum delay}{
	\label{thm:mindelay}
	When $\{T_k\}\in \mathbb{DT}_M$ is a full code-tree set, among every $\{T'_k\}\sim \{T_k\}$, $\{\tilde{T}_k\}$ given by $\{T_k\}$ using {\it Procedure \ref{prc:basic}} constructs a code with the shortest decoding delay. 
}
Fig.~\ref{fig:reduced_ex} gives an example. The code-tree set in (a) represents a code for $\mathbb{A}_3$. 
Note that in $T_0$, we have both $a$ and $b$ assigned to the root, and the decodability is guaranteed by using different code trees after encoding $a$ and $b$. 
The code $V(s)$ represented by Fig.~\ref{fig:reduced_ex} (a) gives $V(aa)=$`000', $V(ab)=$`001', $V(ac)=$`010', $V(ba)=$`011', $V(bb)=$`100', $V(bc)=$`101', and $V(c)=$`11'. 
$\{T_k\}$ follows {\it Rule \ref{rle:decodable}} \ref{srle:prefix_in_mode} but there are some useless binary strings in its modes: 
The common prefix `$0$' in $\textsc{Mode}_1$ can be moved to $T_0$ because the encoder at $T_0$ can immediately determine as its output when encoding $a$; 
the decoder at $T_0$ needs only to read `00', instead of `000' or `001' in $\textsc{Mode}_1$, to determine $a$ as the output because any codeword starting with `00' corresponds to symbol sequence starting with $a$; 
`10', instead of `100' or `101' in $\textsc{Mode}_2$, is enough for the decoder at $T_0$ to determine $b$ as the output. 

{\it Procedure \ref{prc:basic}} reduces such useless modes by using only the basic modes. 
Fig.~\ref{fig:reduced_ex} (b) describes the code-tree set given by converting $\{T_k\}$ of the above example. 
The code $\tilde{V}(s)$ represented by Fig.~\ref{fig:reduced_ex} (b) outputs the same codewords as $V(s)$ for $aa$, $ab$, $ac$, $ba$, $bb$, $bc$, and $c$. 
$V(s)$ and $\tilde{V}(s)$ differ only when we use non-zero-length termination codewords. 
For example, $V(ca)=$`11000' comprises a codeword `$11\lambda$' for $ca$ and a termination codeword `$000$' ($\in\textsc{Mode}_1$), 
and $\tilde{V}(ca)=$`1100' comprises a codeword `$110$' for $ca$ and a termination codeword `$0$' ($\in\tilde{\textsc{Mode}}_1$). 
The difference of the outputs lies only in their ends, and we can make $V(ca)$ and $\tilde{V}(ca)$ identical by truncating the termination codeword `$000$'. 
The binary strings in the modes get shortened by the conversion with a trivial change in the output codewords. 

We use the following lemma to prove the above theorems. 
\mytheory{lmm}{Properties of the reduced binary string set}{
	\label{lmm:redset}	
	$F_{\rm red}(\textsc{Words})$ obeys the following properties for any $\textsc{Words}\in\mathbb{M}$. 
	\begin{enumerate}
		\item $\forall \hat{w}\in F_{\rm red}(\textsc{Words}), \exists w\in \textsc{Words}: \hat{w}\preceq w$.
		\label{slmm:always_be_prefix}
		\item $\forall w\in \textsc{Words}, \exists \hat{w}\in F_{\rm red}(\textsc{Words}): \hat{w}\preceq w$.
		\label{slmm:always_have_prefix}
		\item $(\forall w\in \textsc{Words}, \exists w'\in \textsc{Words}': w'\preceq w) \Longrightarrow(\forall \hat{w}\in F_{\rm red}(\textsc{Words}), \exists \hat{w}'\in F_{\rm red}(\textsc{Words}'): \hat{w}'\preceq \hat{w})$.
		\label{slmm:preserve_having_prefix}
		\item $\forall \hat{w}\in F_{\rm red}(\textsc{Words}), \forall w'\in\mathbb{W}: (\hat{w}\parallel w' \Longrightarrow (\exists w\in \textsc{Words}: w'\parallel w))$. 
		\label{slmm:path_check}
		\item $\forall\textrm{Prefix}\in \mathbb{W}: F_{\rm red}(\{\textrm{Prefix}\:w\mid w\in\textsc{Words}\}) = \{\textrm{Prefix}\:\hat{w}\mid \hat{w}\in F_{\rm red}(\textsc{Words})\}$. 
		\label{slmm:preserve_cmn_prefix}
		\item $F_{\rm red}(F_{\rm red}(\textsc{Words}))=F_{\rm red}(\textsc{Words})$. 
		\label{slmm:irreducible}
	\end{enumerate}
}
Propositions \ref{slmm:always_be_prefix} and \ref{slmm:always_have_prefix} show that $F_{\rm red}$ simply shortens the strings and never extends or neglects them. 
It is guaranteed by \ref{slmm:preserve_having_prefix} that $F_{\rm red}$ preserves the relationship of $\textsc{Words}$ always being some prefix of $\textsc{Words}'$. 
Owing to \ref{slmm:path_check}, we can check whether some string $w$ is related to $\textsc{Words}$ by checking the relation between $w$ and $F_{\rm red}(\textsc{Words})$. 
Additionally, according to \ref{slmm:preserve_cmn_prefix} and \ref{slmm:irreducible}, $F_{\rm red}$ also preserves common prefixes and outputs irreducible binary string sets. 
\\\\
\underline{{\it Proof of Lemma \ref{lmm:redset}}}: The following can be said for any $\textsc{Words}\in\mathbb{M}$. 

{\bf \ref{slmm:always_be_prefix}}. $F_{\rm red}(\textsc{Words})\subseteq F_{\rm full}(\textsc{Words})$ from the definition of $F_{\rm red}$. 
Assume there is some string $\hat{w} \in F_{\rm red}(\textsc{Words})$ that is $\hat{w}\npreceq w$ for any $w\in \textsc{Words}$. 
Combining the fact with the definition of $F_{\rm full}$, we have $\forall \textrm{Suffix}\in\mathbb{W}, \exists w\in \textsc{Words}: w\prec \hat{w}\:\textrm{Suffix}$. 
However, since $w\in \textsc{Words}\subseteq F_{\rm full}(\textsc{Words})$ is obvious from the definition of $F_{\rm full}$, $w\prec \hat{w}$ conflicts with $\hat{w}\in F_{\rm red}(\textsc{Words})$. 
Therefore, every $\hat{w} \in F_{\rm red}(\textsc{Words})$ satisfies $\exists w\in \textsc{Words}:\hat{w}\preceq w$. \\

{\bf \ref{slmm:always_have_prefix}}. If we assume some $w\in \textsc{Words}$ that meets $\forall \hat{w}\in F_{\rm red}(\textsc{Words}): \hat{w}\npreceq w$, it is naturally $w\notin F_{\rm red}(\textsc{Words})$. 
Since $w\in \textsc{Words}\subseteq F_{\rm full}(\textsc{Words})$ and $w\notin F_{\rm red}(\textsc{Words})$, 
we can derive from the definition of $F_{\rm red}$ that there must be some $\hat{w}\in F_{\rm red}(\textsc{Words})$ satisfying $\hat{w}\prec w$. 
However, this conflicts with the assumption, and thus every $w\in \textsc{Words}$ meets $\exists \hat{w}\in F_{\rm red}(\textsc{Words}): \hat{w}\preceq w$. \\

{\bf \ref{slmm:preserve_having_prefix}}. Assume $\hat{w}\in F_{\rm red}(\textsc{Words})$ satisfying $\forall \hat{w}'\in F_{\rm red}(\textsc{Words}'): \hat{w}' \npreceq \hat{w}$ when $\forall w\in \textsc{Words}, \exists w'\in \textsc{Words}': w'\preceq w$. 
The assumption naturally gives $\forall \hat{w}'\in F_{\rm red}(\textsc{Words}'): \hat{w}' \nprec \hat{w}$ and $\hat{w}\notin F_{\rm red}(\textsc{Words}')$, which leads to $\hat{w}\notin F_{\rm full}(\textsc{Words}')$. 
Since $\hat{w}\in F_{\rm red}(\textsc{Words})$, it meets $\forall \textrm{Suffix}\in\mathbb{W}, \exists w\in \textsc{Words}: \hat{w}\:\textrm{Suffix}\parallel w$. 
However, combining it with the assumption, we have $\forall \textrm{Suffix}\in\mathbb{W}, \exists w'\in \textsc{Words}': \hat{w}\:\textrm{Suffix}\parallel w'$, which conflicts with $\hat{w}\notin F_{\rm full}(\textsc{Words}')$. 
Therefore, $\exists \hat{w}'\in F_{\rm red}(\textsc{Words}'): \hat{w}' \preceq \hat{w}$ must always hold. \\

{\bf \ref{slmm:path_check}}. For any $\hat{w}\in F_{\rm red}(\textsc{Words})$ and $w'\in \mathbb{W}$, it is obvious from $\hat{w}\in F_{\rm red}(\textsc{Words})$ that $\exists w\in \textsc{Words}: w'\parallel w$ if $w'\prec \hat{w}$. 
If $\hat{w}\preceq w'$, we can make $\hat{w}\:\textrm{Suffix}=w'$ by some $\textrm{Suffix}\in \mathbb{W}$. 
From the definition of $F_{\rm full}$, we have $\exists w\in \textsc{Words}: \hat{w}\:\textrm{Suffix}\parallel w$, and thus there is always some $w$ that satisfies $w'\parallel w$. \\

{\bf \ref{slmm:preserve_cmn_prefix}}. For $\textrm{Prefix}\in \mathbb{W}$, let us write as $\textsc{Words}'\equiv \{\textrm{Prefix}\:w\mid w\in\textsc{Words}\}$ and $\hat{\textsc{Words}}'\equiv \{\textrm{Prefix}\:\hat{w}\mid \hat{w}\in F_{\rm red}(\textsc{Words})\}$. 
If $\hat{w}\in F_{\rm red}(\textsc{Words})$, it satisfies $\forall \textrm{Suffix}\in \mathbb{W}, \exists w\in \textsc{Words}: \hat{w}\:\textrm{Suffix}\parallel w$. 
Equivalently, it is $\forall \textrm{Suffix}\in \mathbb{W}, \exists \textrm{Prefix}\:w\in \textsc{Words}': \textrm{Prefix}\:\hat{w}\:\textrm{Suffix}\parallel \textrm{Prefix}\:w$. 
Therefore, $\textrm{Prefix}\:\hat{w}\in F_{\rm red}(\textsc{Words}')$, namely $\hat{\textsc{Words}}'\subseteq F_{\rm red}(\textsc{Words}')$. 

On the other hand, if $\hat{w}'\in F_{\rm red}(\textsc{Words}')$, we have 
\begin{equation}
	\forall \textrm{Suffix}\in \mathbb{W}, \exists \textrm{Prefix}\:w\in \textsc{Words}': \hat{w}'\:\textrm{Suffix}\parallel \textrm{Prefix}\:w.
	\label{eq:w_hat_dash}
\end{equation} 
It is $\textrm{Prefix}\preceq \hat{w}'$ because if not, Eq.~(\ref{eq:w_hat_dash}) gives $\hat{w}'\prec \textrm{Prefix}$ but becomes false for some $\textrm{Suffix}$ meeting $\textrm{Suffix}\nparallel\hat{w}'\oslash\textrm{Prefix}$. 
Consequently, it can be written as $\hat{w}'=\textrm{Prefix}\:\hat{w}$ with some $\hat{w}\in \mathbb{W}$, and $\forall \textrm{Suffix}\in \mathbb{W}, \exists \textrm{Prefix}\:w\in \textsc{Words}':\textrm{Prefix}\:\hat{w}\:\textrm{Suffix}\parallel \textrm{Prefix}\:w$ holds, which gives $\forall \textrm{Suffix}\in \mathbb{W}, \exists w\in \textsc{Words}:\hat{w}\:\textrm{Suffix}\parallel w$, namely $\hat{w}\in F_{\rm red}(\textsc{Words})$. 
So, we have $F_{\rm red}(\textsc{Words}')\subseteq\hat{\textsc{Words}}'$, and thus $F_{\rm red}(\textsc{Words}')=\hat{\textsc{Words}}'$.\\

{\bf \ref{slmm:irreducible}}. If $\textrm{Prefix}\in F_{\rm full}(F_{\rm red}(\textsc{Words}))$, it meets $\forall \textrm{Suffix}\in \mathbb{W}, \exists \hat{w}\in F_{\rm red}(\textsc{Words}): \textrm{Prefix}\:\textrm{Suffix}\parallel \hat{w}$. 
Using proposition \ref{slmm:path_check}, it becomes $\forall \textrm{Suffix}\in \mathbb{W}, \exists w\in \textsc{Words}: \textrm{Prefix}\:\textrm{Suffix}\parallel w$, and therefore $F_{\rm full}(F_{\rm red}(\textsc{Words})) \subseteq F_{\rm full}(\textsc{Words})$. 

On the other hand, if $\textrm{Prefix}\in F_{\rm full}(\textsc{Words})$, it meets $\forall \textrm{Suffix}\in \mathbb{W}, \exists w\in \textsc{Words}: \textrm{Prefix}\:\textrm{Suffix}\parallel w$. 
Combining it with proposition \ref{slmm:always_have_prefix}, we can get $\forall \textrm{Suffix}\in \mathbb{W}, \exists \hat{w}\in F_{\rm red}(\textsc{Words}): \textrm{Prefix}\:\textrm{Suffix}\parallel \hat{w}$, and $F_{\rm full}(\textsc{Words}) \subseteq F_{\rm full}(F_{\rm red}(\textsc{Words}))$. 
Therefore, $F_{\rm full}(\textsc{Words}) = F_{\rm full}(F_{\rm red}(\textsc{Words}))$. 
Accordingly, 
\begin{eqnarray}
	F_{\rm red}(F_{\rm red}(\textsc{Words}))&=&\{\hat{w}\in F_{\rm full}(\textsc{Words})\mid \forall \textrm{Prefix}\in F_{\rm full}(\textsc{Words}): \textrm{Prefix}\nprec \hat{w}\}\nonumber\\
	&=& F_{\rm red}(\textsc{Words}). \qquad \blacksquare
\end{eqnarray}\\
\underline{{\it Proof of Theorem \ref{thm:equiv}}}: 
We prove the theorem by taking some steps revealing the following.
\begin{enumerate}
	\item It is always possible to make Eq.~(\ref{eq:Ttilde}).  
	\item $\{\tilde{T}_k\}$ satisfies {\it Rule \ref{rle:decodable}} \ref{srle:prefix_free_expansion}. 
	\item $\{\tilde{T}_k\}$ satisfies {\it Rule \ref{rle:decodable}} \ref{srle:prefix_in_mode}. 
	\item If $\textsc{Mode}_0=\{$`$\lambda$'$\}$, we can make the codeword sequences of $\{T_k\}$ and $\{\tilde{T}_k\}$ identical by truncating their termination codewords. 
\end{enumerate}
\vspace{5mm}

{\bf a}. To make Eq.~(\ref{eq:Ttilde}), every string in $F_{\rm red}(\textsc{Mode}_k)$ must have a prefix $f_{\rm cmn}(\textsc{Mode}_k)$, and every $\textrm{Cword}_k(a)f_{\rm cmn}(\textsc{Mode}_{\textrm{Point}_k(a)})$ must have a prefix $f_{\rm cmn}(\textsc{Mode}_k)$. 
Due to {\it Lemma \ref{lmm:redset}} \ref{slmm:preserve_cmn_prefix}, every string in $F_{\rm red}(\textsc{Mode}_k)$ has a prefix $f_{\rm cmn}(\textsc{Mode}_k)$. 

On the other hand, from {\it Rule \ref{rle:decodable}} \ref{srle:prefix_in_mode} for $\{T_k\}$, 
\begin{equation}
	\exists \textrm{Query}_k\in \textsc{Mode}_k: f_{\rm cmn}(\textsc{Mode}_k) \preceq \textrm{Query}_k \preceq \textrm{Cword}_k(a)\textrm{Query}_{\textrm{Point}_k(a)}
	\label{eq:cmn_pre_of_expcd}
\end{equation}
holds for any $\textrm{Query}_{\textrm{Point}_k(a)}\in \textsc{Mode}_{\textrm{Point}_k(a)}$ and $a\in\mathbb{A}_M$. 
Eq.~(\ref{eq:cmn_pre_of_expcd}) implies that $f_{\rm cmn}(\textsc{Mode}_k)$ is a common prefix of $\{\textrm{Cword}_k(a)\textrm{Query}_{\textrm{Point}_k(a)}\mid \textrm{Query}_{\textrm{Point}_k(a)}\in \textsc{Mode}_{\textrm{Point}_k(a)}\}$. 
Since the maximum-length common prefix can be written as $\textrm{Cword}_k(a)f_{\rm cmn}(\textrm{Query}_{\textrm{Point}_k(a)})$, 
every $\textrm{Cword}_k(a)f_{\rm cmn}(\textsc{Mode}_{\textrm{Point}_k(a)})$ has a prefix $f_{\rm cmn}(\textsc{Mode}_k)$. 
So, we can always make $\{\tilde{T}_k\}$ from $\{T_k\}$. \\

{\bf b}. The expanded codeword sets of the code trees in $\{\tilde{T}_k\}$ are written as 
\begin{eqnarray}
	\tilde{\textsc{Expand}}_k(a)&\equiv& \{\tilde{\textrm{Cword}}_k(a)\tilde{\textrm{Query}}_{\textrm{Point}_k(a)}\mid \tilde{\textrm{Query}}_{\textrm{Point}_k(a)}\in\tilde{\textsc{Mode}}_{\textrm{Point}_k(a)}\}\nonumber\\
	&=&\{f_{\rm cmn}(\textsc{Mode}_k)\oslash\textrm{Cword}_k(a)\hat{\textrm{Query}}_{\textrm{Point}_k(a)}\mid \hat{\textrm{Query}}_{\textrm{Point}_k(a)}\in F_{\rm red}(\textsc{Mode}_{\textrm{Point}_k(a)})\}. 
\end{eqnarray}
{\it Lemma \ref{lmm:redset}} \ref{slmm:path_check} can be rewritten using the contraposition as  
\begin{equation}
	\forall \hat{w}\in F_{\rm red}(\textsc{Words}), \forall w'\in\mathbb{W}: ((\forall w\in \textsc{Words}: w'\nparallel w) \Longrightarrow \hat{w}\nparallel w' ). 
	\label{eq:lemma_1_c_alt}
\end{equation}
Substituting $\textsc{Words}=\textsc{Expand}_k(a)$ to Eq.~(\ref{eq:lemma_1_c_alt}) and combining it with $\textsc{Expand}_k(a)\nparallel \textsc{Expand}_k(a')$, from {\it Rule \ref{rle:decodable}} \ref{srle:prefix_free_expansion}, gives $F_{\rm red}(\textsc{Expand}_k(a))\nparallel\textsc{Expand}_k(a')$. 
Applying Eq.~(\ref{eq:lemma_1_c_alt}) again to it with $\textsc{Words}=\textsc{Expand}_k(a')$, we have 
\begin{equation}
	F_{\rm red}(\textsc{Expand}_k(a))\nparallel F_{\rm red}(\textsc{Expand}_k(a')) 
\end{equation}
for any $a\neq a'$.
Based on {\it Lemma \ref{lmm:redset}} \ref{slmm:preserve_cmn_prefix}, 
\begin{eqnarray}
	F_{\rm red}(\textsc{Expand}_k(a)) &=&  F_{\rm red}(\{\textrm{Cword}_k(a)\textrm{Query}_{\textrm{Point}_k(a)}\mid \textrm{Query}_{\textrm{Point}_k(a)}\in\textsc{Mode}_{\textrm{Point}_k(a)}\})\nonumber\\
	&=& \{\textrm{Cword}_k(a)\hat{\textrm{Query}}_{\textrm{Point}_k(a)}\mid \hat{\textrm{Query}}_{\textrm{Point}_k(a)}\in F_{\rm red}(\textsc{Mode}_{\textrm{Point}_k(a)})\},
\end{eqnarray}
and therefore
\begin{eqnarray}
	&&\{\textrm{Cword}_k(a)\hat{\textrm{Query}}_{\textrm{Point}_k(a)}\mid \hat{\textrm{Query}}_{\textrm{Point}_k(a)}\in F_{\rm red}(\textsc{Mode}_{\textrm{Point}_k(a)})\} \nonumber\\
	&\nparallel& \{\textrm{Cword}_k(a')\hat{\textrm{Query}}_{\textrm{Point}_k(a')}\mid \hat{\textrm{Query}}_{\textrm{Point}_k(a')}\in F_{\rm red}(\textsc{Mode}_{\textrm{Point}_k(a')})\}. 
	\label{eq:expc_red_npara}
\end{eqnarray}
Since $f_{\rm cmn}(\textsc{Mode}_k) \preceq \textrm{Cword}_k(a)\hat{\textrm{Query}}_{\textrm{Point}_k(a)}$ and $f_{\rm cmn}(\textsc{Mode}_k) \preceq \textrm{Cword}_k(a')\hat{\textrm{Query}}_{\textrm{Point}_k(a')}$ from Eq.~(\ref{eq:cmn_pre_of_expcd}), we can derive from Eq.~(\ref{eq:expc_red_npara}) that 
\begin{equation}
	\tilde{\textsc{Expand}}_k(a) \nparallel \tilde{\textsc{Expand}}_k(a'). 
\end{equation}
Therefore, $\{\tilde{T}_k\}$ obeys {\it Rule \ref{rle:decodable}} \ref{srle:prefix_free_expansion}. \\

{\bf c}. From the definition of $\textsc{Expands}_k$, we can write {\it Rule \ref{rle:decodable}} \ref{srle:prefix_in_mode} as
\begin{equation}
	\forall \textrm{Query}_{\textrm{Point}_k(a)}\in \textsc{Mode}_{\textrm{Point}_k(a)}, \exists \textrm{Query}_k\in \textsc{Mode}_k: \textrm{Query}_k\preceq \textrm{Cword}_k(a)\textrm{Query}_{\textrm{Point}_k(a)}
	\label{eq:rule_3_b_alt}
\end{equation}
for all $a\in\mathbb{A}_M$. 
Using {\it Lemma \ref{lmm:redset}} \ref{slmm:preserve_having_prefix} and \ref{slmm:preserve_cmn_prefix} to Eq.~(\ref{eq:rule_3_b_alt}), we get
\begin{equation}
	\forall \hat{\textrm{Query}}_{\textrm{Point}_k(a)}\in F_{\rm red}(\textsc{Mode}_{\textrm{Point}_k(a)}), \exists \hat{\textrm{Query}}_k\in F_{\rm red}(\textsc{Mode}_k): \hat{\textrm{Query}}_k\preceq \textrm{Cword}_k(a)\hat{\textrm{Query}}_{\textrm{Point}_k(a)}
\end{equation}
for all $a$. Therefore, 
\begin{equation}
	\forall \tilde{\textrm{Expcw}}\in \tilde{\textsc{Expand}}_k(a), \exists \tilde{\textrm{Query}}_k\in \tilde{\textsc{Mode}}_k: \tilde{\textrm{Query}}_k\preceq \tilde{\textrm{Expcw}}
	\label{eq:rule_3_b_Ttilde}
\end{equation}
so that $\{\tilde{T}_k\}$ satisfies {\it Rule \ref{rle:decodable}} \ref{srle:prefix_in_mode} as well. \\

{\bf d}. For any source symbol sequence $x_0x_1\cdots x_{L-1}$, the encoder using the code-tree set $\{T_k\}$ gives a codeword sequence as 
\begin{eqnarray}
	&&\textrm{Cword}_{k_0}(x_0)\textrm{Cword}_{k_1}(x_1)\cdots \textrm{Cword}_{k_{L-1}}(x_{L-1})\textrm{Query}_{k_L}\nonumber\\
	&=&\textrm{Cword}_{k_0}(x_0)\textrm{Cword}_{k_1}(x_1)\cdots \textrm{Cword}_{k_{L-1}}(x_{L-1})f_{\rm cmn}(\textsc{Mode}_{k_{L}})(f_{\rm cmn}(\textsc{Mode}_{k_{L}})\oslash \textrm{Query}_{k_L})
	\label{eq:T_code}
\end{eqnarray} 
where $k_0=0$ and $k_{i+1}=\textrm{Point}_{k_i}(x_i)$ ($i\in \mathbb{Z}^+_{<L+1}$) are the indexes of the code trees used for the encoding, and $\textrm{Query}_{k_L}\in \textsc{Mode}_{k_L}$ is the termination codeword. 
Meanwhile, when $f_{\rm cmn}(\textsc{Mode}_{k_0})=$`$\lambda$', the codeword sequence given by using $\{\tilde{T_k}\}$ is 
\begin{eqnarray}
	&&(f_{\rm cmn}(\textsc{Mode}_{k_0})\oslash \textrm{Cword}_{k_0}(x_0)f_{\rm cmn}(\textsc{Mode}_{k_1}))(f_{\rm cmn}(\textsc{Mode}_{k_1})\oslash \textrm{Cword}_{k_1}(x_1)f_{\rm cmn}(\textsc{Mode}_{k_2}))\cdots \nonumber\\
	&&\:\:\:\:(f_{\rm cmn}(\textsc{Mode}_{L-1})\oslash \textrm{Cword}_{k_{L-1}}(x_{L-1})f_{\rm cmn}(\textsc{Mode}_{k_{L}}))(f_{\rm cmn}(\textsc{Mode}_{k_{L}})\oslash \hat{\textrm{Query}}_{k_L}) \nonumber\\
	&=& \textrm{Cword}_{k_0}(x_0)\textrm{Cword}_{k_1}(x_1)\cdots \textrm{Cword}_{k_{L-1}}(x_{L-1})f_{\rm cmn}(\textsc{Mode}_{k_{L}})(f_{\rm cmn}(\textsc{Mode}_{k_{L}})\oslash \hat{\textrm{Query}}_{k_L})
	\label{eq:Ttilde_code}
\end{eqnarray} 
where $f_{\rm cmn}(\textsc{Mode}_{k_{L}})\oslash \hat{\textrm{Query}}_{k_L} \in \tilde{\textsc{Mode}}_{k_L}$ is the termination codeword. 
We can make Eqs.~(\ref{eq:T_code}) and (\ref{eq:Ttilde_code}) identical by truncating their termination codewords. \\

It is $\{\tilde{T}_k\}\in\mathbb{DT}_M$ from steps b and c. 
Therefore, step d leads to $\{\tilde{T}_k\}\sim \{T_k\}$. 
$\qquad \blacksquare$\\\\
\underline{{\it Proof of Theorem \ref{thm:mindelay}}}: 
As the definition, we have $f_{\rm cmn}(\textsc{Mode}_0)=$`$\lambda$' when $\{T_k\}$ is a full code-tree set. 
If any $f_{\rm cmn}(\textsc{Mode}_k)$ is not `$\lambda$', thinking similarly to Eq.~(\ref{eq:Ttilde_code}) in the proof of {\it Theorem \ref{thm:equiv}}, 
we can obviously shorten the decoding delay from $\{T_k\}$ by altering $\textsc{Mode}_k$ and $\textrm{Cword}_k(a)$ respectively as $\{f_{\rm cmn}(\textsc{Mode}_k)\oslash\textrm{Query}_k\mid \textrm{Query}_k\in \textsc{Mode}_k\}$ and $f_{\rm cmn}(\textsc{Mode}_k)\oslash \textrm{Cword}_k(a)f_{\rm cmn}(\textsc{Mode}_{\textrm{Point}_k(a)})$. 
Therefore, without loss of generality, we here only discuss the cases of 
\begin{equation}
	\forall k:f_{\rm cmn}(\textsc{Mode}_k)=\text{`$\lambda$'}. 
	\label{eq:assump_no_cmn_prefix}
\end{equation}
In such cases, we can rewrite $\{\tilde{T}_k\}$ as
\begin{equation}
	\label{eq:Ttilde_alt}
	\{\tilde{T}_k = (\textrm{Cword}_k, \textrm{Point}_k, \tilde{\textsc{Mode}}_k, k)\},
\end{equation}	
where
\begin{equation}
	\label{eq:conv1_alt}
	\tilde{\textsc{Mode}}_k = F_{\rm red}(\textsc{Mode}_k).
\end{equation}
The expanded codeword sets become 
\begin{equation}
	\tilde{\textsc{Expand}}_k(a)= \{\textrm{Cword}_k(a)\tilde{\textrm{Query}}_{\textrm{Point}_k(a)}\mid \tilde{\textrm{Query}}_{\textrm{Point}_k(a)}\in F_{\rm red}(\textsc{Mode}_{\textrm{Point}_k(a)})\},
\end{equation}
and say $\tilde{\textsc{Expands}}_k\equiv\{\tilde{\textsc{Expand}}_k(a)\mid a\in\mathbb{A}_M\}$. 

Since $\{T_k\}$ is a full code-tree set, $F_{\rm red}(\textsc{Mode}_k) = F_{\rm red}(\textsc{Expands}_k)$, 
which becomes
\begin{equation}
	\forall \textrm{Query}_k \in \textsc{Mode}_k, \exists \hat{\textrm{Expcw}}\in F_{\rm red}(\textsc{Expands}_k)\text{: }\textrm{Query}_k \preceq \hat{\textrm{Expcw}}
\end{equation}
using {\it Lemma \ref{lmm:redset}} \ref{slmm:always_have_prefix} and 
\begin{equation}
	\forall \textrm{Query}_k \in \textsc{Mode}_k, \exists \textrm{Expcw}\in \textsc{Expands}_k\text{: }\textrm{Query}_k \preceq \textrm{Expcw}
	\label{eq:assum_no_invalid_query}
\end{equation}
using {\it Lemma \ref{lmm:redset}} \ref{slmm:always_be_prefix}.
It means that every binary string in the modes is a prefix of some expanded codeword. 

Let us write a codeword set $\{T'_k\}$ ($\in\mathbb{DT}_M$,  $\sim \{T_k\}$) as 
\begin{eqnarray}
	\{T'_k = (\textrm{Cword}'_k, \textrm{Point}'_k, \textsc{Mode}'_k, k)\}  
\end{eqnarray}
and its expanded codeword sets as
\begin{eqnarray}
	\textsc{Expand}'_k(a)&\equiv& \{\textrm{Cword}'_k(a)\textrm{Query}'_{\textrm{Point}'_k(a)}\mid \textrm{Query}'_{\textrm{Point}'_k(a)}\in \textsc{Mode}'_{\textrm{Point}'_k(a)}\}\\
	\textsc{Expands}'_k&\equiv&\{\textsc{Expand}'_k(a)\mid a\in\mathbb{A}_M\}.
\end{eqnarray}
Since $\{T'_k\}\sim \{T_k\}$, it must be $f_{\rm cmn}(\textsc{Mode}'_0)=$`$\lambda$'. 
For the same reason above, we can assume  
\begin{equation}
	\forall k:f_{\rm cmn}(\textsc{Mode}'_k)=\text{`$\lambda$'}
	\label{eq:assump_no_cmn_prefix_Tdash}
\end{equation}
without loss of generality. 
Still without loss of generality, we can also assume 
\begin{equation}
	\forall \textrm{Query}'_k \in \textsc{Mode}'_k, \exists \textrm{Expcw}'\in \textsc{Expands}'_k\text{: }\textrm{Query}'_k \preceq \textrm{Expcw}'.
	\label{eq:assum_no_invalid_query_Tdash}
\end{equation}
This is because we can just omit $\textrm{Query}'_k$ from $\textsc{Mode}'_k$, without affecting the decodability and decoding delay, if it is not a prefix of any expanded codeword. 

For any source symbol sequence $x_0x_1\cdots x_{L-1}$, the encoder using the code-tree set $\{T'_k\}$ gives a codeword sequence as 
\begin{equation}
	\textrm{Cword}'_{k'_0}(x_0)\textrm{Cword}'_{k'_1}(x_1)\cdots \textrm{Cword}'_{k'_{L-1}}(x_{L-1})\textrm{Query}'_{k'_L}
\end{equation} 
where $k'_0=0$ and $k'_{i+1}=\textrm{Point}'_{k'_i}(x_i)$ ($i\in \mathbb{Z}^+_{<L+1}$) are the indexes of the code trees used for the encoding, and $\textrm{Query}'_{k'_L}\in \textsc{Mode}'_{k'_L}$ is the termination codeword. 

We prove the theorem by taking some steps showing the following facts. 
\begin{enumerate}
	\item $\{\tilde{T}_k\}$ is a full code-tree set. 
	\item If there is some $k'$ for each $k$ that meets $\forall \tilde{\textrm{Query}}_k \in \tilde{\textsc{Mode}}_k, \exists \textrm{Query}'_{k'}\in \textsc{Mode}'_{k'}: \|\tilde{\textrm{Query}}_k\|_{\rm len}\le \|\textrm{Query}'_{k'}\|_{\rm len}$, the decoding delay using $\{\tilde{T}_k\}$ is not longer than the one using $\{T'_k\}$. 
	\item For any $x_0x_1\cdots x_{L-1}$ and $L$, there is some $w\in\mathbb{W}$ that makes $\forall \tilde{\textrm{Query}}_{k_{i}}\in\tilde{\textsc{Mode}}_{k_{i}}, \exists \textrm{Query}'_{k'_{i}}\in\textsc{Mode}'_{k'_{i}}: w\:\tilde{\textrm{Query}}_{k_{i}}\parallel \textrm{Query}'_{k'_{i}}$. 
	\item The condition in step c becomes $w\:\tilde{\textrm{Query}}_{k_{i}}\preceq \textrm{Query}'_{k'_{i}}$ when $\{\tilde{T}_k\}$ is a full code-tree set. 
\end{enumerate}
\vspace{5mm}

{\bf a}. Since $\{T_k\}$ is a full code-tree set, $F_{\rm red}(\textsc{Mode}_k) = F_{\rm red}(\textsc{Expands}_k)$. 
Using {\it Lemma \ref{lmm:redset}} \ref{slmm:irreducible}, we have
\begin{equation}
	F_{\rm red}(\tilde{\textsc{Mode}}_k) = F_{\rm red}(F_{\rm red}(\textsc{Mode}_k))
	= F_{\rm red}(\textsc{Mode}_k)
	= F_{\rm red}(\textsc{Expands}_k).
	\label{eq:mode_eq_expcw}
\end{equation}
When $\textrm{Prefix}\in F_{\rm full}(\tilde{\textsc{Expands}}_k)$, it must satisfy 
\begin{eqnarray}
	&&\forall \textrm{Suffix}\in \mathbb{W}, \exists \tilde{\textrm{Expcw}}\in \tilde{\textsc{Expands}}_k: \textrm{Prefix}\:\textrm{Suffix}\parallel \tilde{\textsc{Expands}}_k \nonumber\\
	&\iff& \forall \textrm{Suffix}\in \mathbb{W}, \exists (a\in\mathbb{A}_M, \tilde{\textrm{Query}}_{\textrm{Point}_k(a)}\in F_{\rm red}(\textsc{Mode}_{\textrm{Point}_k(a)})): \textrm{Prefix}\:\textrm{Suffix}\parallel \textrm{Cword}_k(a)\tilde{\textrm{Query}}_{\textrm{Point}_k(a)}.
	\label{eq:full_mode_expcw_tild}
\end{eqnarray}
From {\it Lemma \ref{lmm:redset}} \ref{slmm:path_check}, Eq.~(\ref{eq:full_mode_expcw_tild}) becomes 
\begin{eqnarray}
	&&\forall \textrm{Suffix}\in \mathbb{W}, \exists (a\in\mathbb{A}_M, \textrm{Query}_{\textrm{Point}_k(a)}\in \textsc{Mode}_{\textrm{Point}_k(a)}): \textrm{Prefix}\:\textrm{Suffix}\parallel \textrm{Cword}_k(a)\textrm{Query}_{\textrm{Point}_k(a)}\nonumber\\
	&\iff& \forall \textrm{Suffix}\in \mathbb{W}, \exists \textrm{Expcw}\in \textsc{Expands}_k: \textrm{Prefix}\:\textrm{Suffix}\parallel \textsc{Expands}_k,
\end{eqnarray}
and thus $\textrm{Prefix}\in F_{\rm full}(\textsc{Expands}_k)$, namely $F_{\rm full}(\tilde{\textsc{Expands}}_k)\subseteq F_{\rm full}(\textsc{Expands}_k)$. 

Similarly, when $\textrm{Prefix}\in F_{\rm full}(\textsc{Expands}_k)$, it must satisfy 
\begin{equation}
	\forall \textrm{Suffix}\in \mathbb{W}, \exists (a\in\mathbb{A}_M, \textrm{Query}_{\textrm{Point}_k(a)}\in \textsc{Mode}_{\textrm{Point}_k(a)}): \textrm{Prefix}\:\textrm{Suffix}\parallel \textrm{Cword}_k(a)\textrm{Query}_{\textrm{Point}_k(a)},
\end{equation}
which becomes
\begin{eqnarray}
	&&\forall \textrm{Suffix}\in \mathbb{W}, \exists (a\in\mathbb{A}_M, \tilde{\textrm{Query}}_{\textrm{Point}_k(a)}\in F_{\rm red}(\textsc{Mode}_{\textrm{Point}_k(a)})): \textrm{Prefix}\:\textrm{Suffix}\parallel \textrm{Cword}_k(a)\tilde{\textrm{Query}}_{\textrm{Point}_k(a)}\nonumber\\
	&\iff& \forall \textrm{Suffix}\in \mathbb{W}, \exists \tilde{\textrm{Expcw}}\in \tilde{\textsc{Expands}}_k: \textrm{Prefix}\:\textrm{Suffix}\parallel \tilde{\textsc{Expands}}_k 
\end{eqnarray}
according to {\it Lemma \ref{lmm:redset}} \ref{slmm:always_have_prefix}. 
So, it is also $F_{\rm full}(\textsc{Expands}_k)\subseteq F_{\rm full}(\tilde{\textsc{Expands}}_k)$. 

Based on the fact above, it is $F_{\rm full}(\textsc{Expands}_k)= F_{\rm full}(\tilde{\textsc{Expands}}_k)$ and thus
\begin{eqnarray}
	F_{\rm red}(\tilde{\textsc{Expands}}_k)&=&\{\hat{w}\in F_{\rm full}(\textsc{Expands}_k)\mid \forall \textrm{Prefix}\in F_{\rm full}(\textsc{Expands}_k): \textrm{Prefix}\nprec \hat{w}\}\nonumber\\
	&=& F_{\rm red}(\textsc{Expands}_k).
\end{eqnarray}
With Eq.~(\ref{eq:mode_eq_expcw}), we have $F_{\rm red}(\tilde{\textsc{Mode}}_k)=F_{\rm red}(\tilde{\textsc{Expands}}_k)$. 
Therefore, $\{\tilde{T}_k\}$ is also a full code-tree set. \\

{\bf b}. Assume there is some $\tilde{\textrm{Query}}_k \in \tilde{\textsc{Mode}}_k$ that is $\forall \tilde{\textrm{Expcw}}\in \tilde{\textsc{Expands}}_k\text{: }\tilde{\textrm{Query}}_k \npreceq \tilde{\textrm{Expcw}}$. 
In this case, it has to be either $\tilde{\textrm{Expcw}}\prec \tilde{\textrm{Query}}_k$ or $\tilde{\textrm{Query}}_k \nparallel \tilde{\textrm{Expcw}}$. 
If $\tilde{\textrm{Expcw}}\prec \tilde{\textrm{Query}}_k$, because of Eq.~(\ref{eq:rule_3_b_Ttilde}) derived from {\it Rule \ref{rle:decodable}} \ref{srle:prefix_in_mode} for $\{\tilde{T}_k\}$, there is some $\tilde{\textrm{Query}}^*_k\in \tilde{\textsc{Mode}}_k$ satisfying $\tilde{\textrm{Query}}^*_k\preceq \tilde{\textrm{Expcw}}$. It leads to $\tilde{\textrm{Query}}^*_k\prec \tilde{\textrm{Query}}_k$, which conflicts with $\tilde{\textsc{Mode}}_k\in \mathbb{PF}$. 

Due to this fact, under the assumption above, it must be $\forall \tilde{\textrm{Expcw}}\in \tilde{\textsc{Expands}}_k\text{: }\tilde{\textrm{Query}}_k \nparallel \tilde{\textrm{Expcw}}$, namely $\exists \tilde{\textrm{Query}}_k\in F_{\rm red}(\textsc{Mode}_k), \forall (a\in\mathbb{A}_M, \tilde{\textrm{Query}}_{\textrm{Point}_k(a)}\in F_{\rm red}(\textsc{Mode}_{\textrm{Point}_k(a)})\text{: }\tilde{\textrm{Query}}_k \nparallel \textrm{Cword}_k(a)\tilde{\textrm{Query}}_{\textrm{Point}_k(a)}$. 
However, using {\it Lemma \ref{lmm:redset}} \ref{slmm:always_be_prefix}, it becomes $\exists \textrm{Query}_k\in \textsc{Mode}_k, \forall (a\in\mathbb{A}_M, \textrm{Query}_{\textrm{Point}_k(a)}\in \textsc{Mode}_{\textrm{Point}_k(a)})\text{: }\textrm{Query}_k \nparallel \textrm{Cword}_k(a)\textrm{Query}_{\textrm{Point}_k(a)}$, which conflicts with Eq.~(\ref{eq:assum_no_invalid_query}). 
Therefore, 
\begin{equation}
	\forall \tilde{\textrm{Query}}_k \in \tilde{\textsc{Mode}}_k, \exists \tilde{\textrm{Expcw}}\in \tilde{\textsc{Expands}}_k\text{: }\tilde{\textrm{Query}}_k \preceq \tilde{\textrm{Expcw}}
	\label{eq:no_invalid_query_Ttilde}
\end{equation}
holds. 

Owing to Eqs.~(\ref{eq:assum_no_invalid_query_Tdash}) and (\ref{eq:no_invalid_query_Ttilde}), 
the values of the decoding delay given by Eq.~(\ref{eq:delay}) respectively for $\{\tilde{T}_k\}$ and $\{T'_k\}$ are 
\begin{equation}
	\max_k \left(\max \{\|\tilde{\textrm{Query}}_k\|_{\rm len}\mid\tilde{\textrm{Query}}_k\in \tilde{\textsc{Mode}}_k\}\right)
\end{equation}
\begin{equation}
	\max_{k'} \left(\max \{\|\textrm{Query}'_{k'}\|_{\rm len}\mid\textrm{Query}'_{k'}\in \textsc{Mode}'_{k'}\}\right).
\end{equation}
Thus, at least if there is some $k'$ for each $k$ that meets $\forall \tilde{\textrm{Query}}_k \in \tilde{\textsc{Mode}}_k, \exists \textrm{Query}'_{k'}\in \textsc{Mode}'_{k'}: \|\tilde{\textrm{Query}}_k\|_{\rm len}\le \|\textrm{Query}'_{k'}\|_{\rm len}$, the decoding delay using $\{\tilde{T}_k\}$ is not longer than the one using $\{T'_k\}$. \\

\begin{figure}[tb]
	\begin{center}
		\includegraphics[width=13cm,  bb=0 0 794 228]{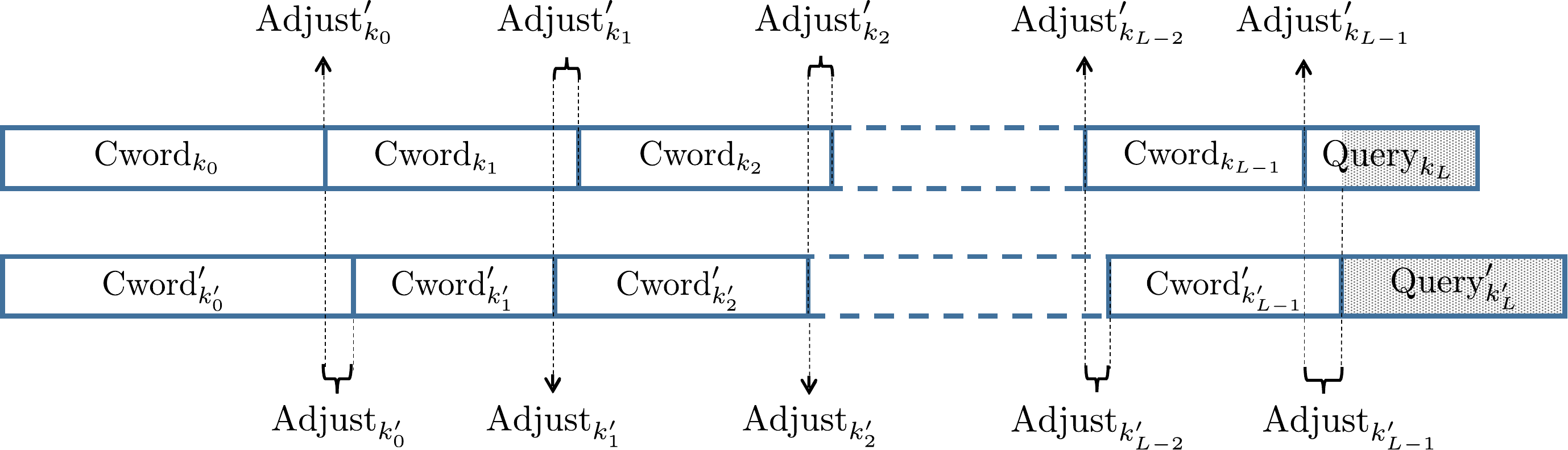}
	\end{center}
	\caption{Example of the relationship represented by Eq.~(\ref{eq:adjust_codeword}) between the codeword sequences given respectively by $\{T_k\}$ and $\{T'_K\}$. $\{T_k\}\sim\{T'_K\}$ requires both codeword sequences to be identical except for the grayed-out area. }
	\label{fig:adjust_ex}
\end{figure}
{\bf c}. Since $\{T'_k\}\sim \{T_k\}$, for any source symbol sequence $x_0x_1\cdots x_{L-1}$, there must be some binary strings $\textrm{Adjust}'_{k_i}(x_i)$ and $\textrm{Adjust}_{k'_i}(x_i)$ ($\in\mathbb{W}$, $i\in\mathbb{Z}^+_{<L}$) giving 
\begin{equation}
	\left \{
	\begin{array}{rcl}
		\textrm{Cword}'_{0}(x_0)\textrm{Adjust}'_{0}(x_0) &=& \textrm{Cword}_{0}(x_0)\textrm{Adjust}_{0}(x_0)\\
		\textrm{Adjust}'_{k_{i-1}}(x_{i-1})\oslash(\textrm{Cword}'_{k'_i}(x_i)\textrm{Adjust}'_{k_i}(x_i)) &=& \textrm{Adjust}_{k'_{i-1}}(x_{i-1})\oslash(\textrm{Cword}_{k_i}(x_i)\textrm{Adjust}_{k'_i}(x_i))
	\end{array}
	\right. ,
	\label{eq:adjust_codeword}
\end{equation}
as illustrated in Fig.~\ref{fig:adjust_ex}. 
If $\textrm{Adjust}_{k'_{i-1}}(x_{i-1})\neq\text{`$\lambda$'}$ ($i>0$) exists, it means $\{\textrm{Cword}_{k_i}(x_i)\mid x_i\in \mathbb{A}_M\}$ has a common prefix of at least 1 bit. 
If so, $f_{\rm cmn}(\textsc{Expands}_{k_i})\neq\text{`$\lambda$'}$, and from {\it Lemma \ref{lmm:redset}} \ref{slmm:preserve_cmn_prefix}, $f_{\rm cmn}(F_{\rm red}(\textsc{Expands}_{k_i}))\neq\text{`$\lambda$'}$. 
Since $\{T_k\}$ is a full code-tree set, it must be $f_{\rm cmn}(F_{\rm red}(\textsc{Mode}_{k_i}))=f_{\rm cmn}(F_{\rm red}(\textsc{Expands}_{k_i}))\neq\text{`$\lambda$'}$, and therefore $f_{\rm cmn}(\textsc{Mode}_{k_i})\neq\text{`$\lambda$'}$ from {\it Lemma \ref{lmm:redset}} \ref{slmm:preserve_cmn_prefix}. 
This fact conflicts with the assumption of Eq.~(\ref{eq:assump_no_cmn_prefix}), and thus $\textrm{Adjust}_{k'_{i-1}}(x_{i-1})$ must be `$\lambda$', resulting in 
\begin{equation}
	\textrm{Cword}_{k_i}(x_i) = \textrm{Adjust}'_{k_{i-1}}(x_{i-1})\oslash(\textrm{Cword}'_{k'_i}(x_i)\textrm{Adjust}'_{k_i}(x_i)),
	\label{eq:Ttilde_Tdash_code_relation}
\end{equation} 
which also holds for $i=0$ if we define $\textrm{Adjust}_{k'_{-1}}(x_{-1})\equiv$ `$\lambda$'. 

On the other hand, recursively applying {\it Rule \ref{rle:decodable}} \ref{srle:prefix_in_mode} for $\{\tilde{T}_k\}$ implies that there is $\tilde{\textrm{Query}}_{k_{i}}\in \tilde{\textsc{Mode}}_{k_i}$ satisfying 
\begin{equation}
	\tilde{\textrm{Query}}_{k_{i}}\preceq \tilde{\textrm{Cword}}_{k_{i}}(x_{i})\tilde{\textrm{Cword}}_{k_{i+1}}(x_{i+1})\cdots \tilde{\textrm{Cword}}_{k_{L-1}}(x_{L-1})\tilde{\textrm{Query}}_{k_L}
\end{equation}
where $\tilde{\textrm{Query}}_{k_L}\in\tilde{\textsc{Mode}}_{k_L}$. 
Due to Eq.~(\ref{eq:Ttilde_Tdash_code_relation}), this condition is equivalent to
\begin{equation}
	\textrm{Adjust}'_{k_{i-1}}(x_{i-1})\tilde{\textrm{Query}}_{k_{i}}\preceq \textrm{Cword}'_{k'_i}(x_i)\textrm{Cword}'_{k'_{i+1}}(x_{i+1})\cdots \textrm{Cword}'_{k'_{L-1}}(x_{L-1})\textrm{Adjust}'_{k_{L-1}}(x_{L-1})\tilde{\textrm{Query}}_{k_L}.
\end{equation}
Similarly, there is $\textrm{Query}'_{k'_{i}}\in \textsc{Mode}'_{k'_i}$ satisfying
\begin{equation}
	\textrm{Query}'_{k'_{i}}\preceq \textrm{Cword}'_{k'_i}(x_i)\textrm{Cword}'_{k'_{i+1}}(x_{i+1})\cdots \textrm{Cword}'_{k'_{L-1}}(x_{L-1})\textrm{Query}'_{k'_L},
\end{equation}
where $\textrm{Query}'_{k'_L}\in\textsc{Mode}'_{k'_L}$.
If $\textrm{Adjust}'_{k_{i-1}}(x_{i-1})\tilde{\textrm{Query}}_{k_{i}}\nparallel \textrm{Query}'_{k'_{i}}$ in this case, both $\textrm{Adjust}'_{k_{i-1}}(x_{i-1})\tilde{\textrm{Query}}_{k_{i}}$ and $\textrm{Query}'_{k'_{i}}$ must have a common prefix $\textrm{Cword}'_{k'_i}(x_i)\textrm{Cword}'_{k'_{i+1}}(x_{i+1})\cdots \textrm{Cword}'_{k'_{L-1}}(x_{L-1})$. 
However, it cannot hold for arbitrary $L$ because the decoding delay is finite. 
Therefore, it is always $\textrm{Adjust}'_{k_{i-1}}(x_{i-1})\tilde{\textrm{Query}}_{k_{i}}\parallel \textrm{Query}'_{k'_{i}}$. 
As a result, we have 
\begin{equation}
	\forall \tilde{\textrm{Query}}_{k_{i}}\in\tilde{\textsc{Mode}}_{k_{i}}, \exists \textrm{Query}'_{k'_i}\in\textsc{Mode}'_{k'_i}: \textrm{Adjust}'_{k_{i-1}}(x_{i-1})\tilde{\textrm{Query}}_{k_{i}}\parallel \textrm{Query}'_{k'_{i}} 
	\label{eq:mode_relation}
\end{equation}
for any $i\in\mathbb{Z}^+$. 
\\

\begin{figure}[tb]
	\begin{center}
		\includegraphics[width=15cm,  bb=0 0 904 362]{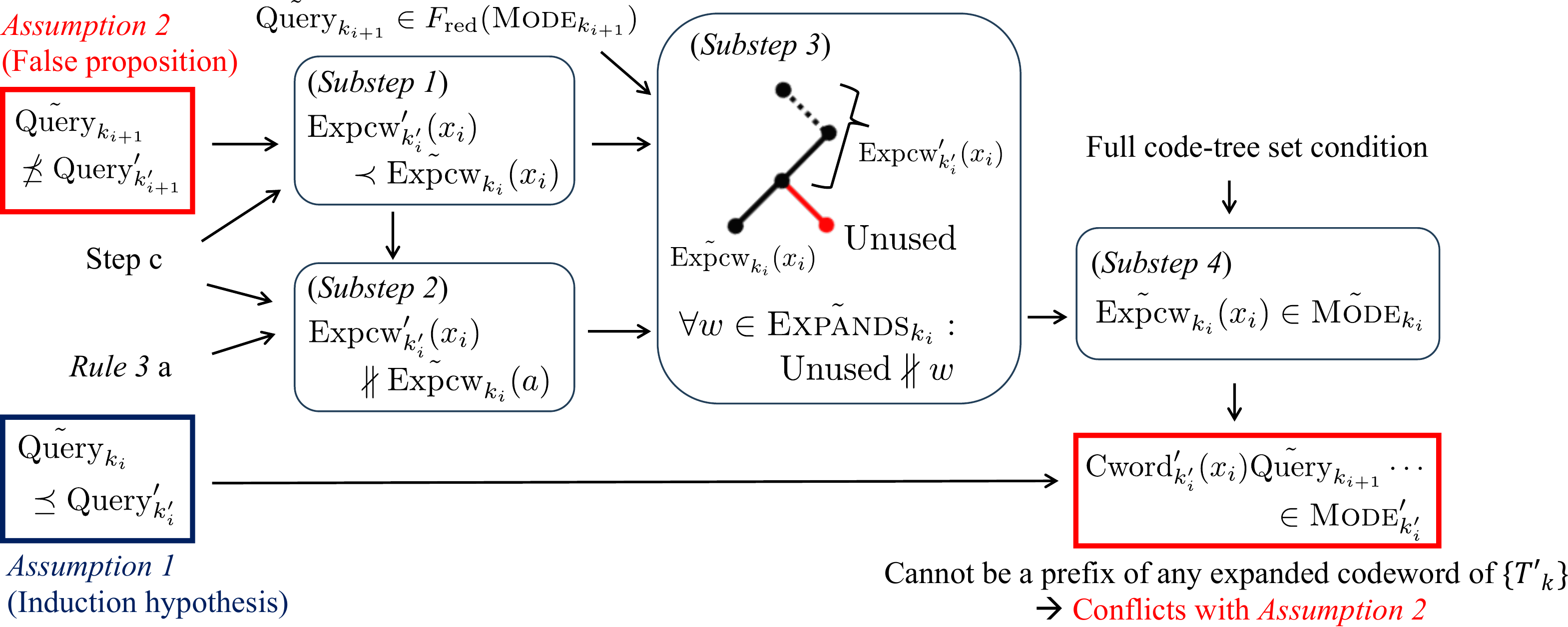}
	\end{center}
	\caption{Outline of step d in the proof of {\it Theorem \ref{thm:mindelay}}. $\textrm{Adjust}'_{k_{i-1}}(x_{i-1})$ and $\textrm{Adjust}'_{k_{i}}(x_{i})$ are omitted here for simplicity.}
	\label{fig:chart_step_d}
\end{figure}

{\bf d}. We show by an inductive approach that it is actually $\textrm{Adjust}'_{k_{i-1}}(x_{i-1})\tilde{\textrm{Query}}_{k_{i}}\preceq \textrm{Query}'_{k'_{i}}$ in Eq.~(\ref{eq:mode_relation}). 
The outline of the proof is depicted in Fig.~\ref{fig:chart_step_d}. 

i) [Base case] Since $\{\tilde{T}_k\}$ is a full code-tree set, $\tilde{\textsc{Mode}}_{k_0}=\{$`$\lambda$'$\}$, and therefore it is always $\textrm{Adjust}'_{k_{-1}}(x_{-1})\tilde{\textrm{Query}}_{k_0}=\tilde{\textrm{Query}}_0=$ `$\lambda$' $\preceq \textrm{Query}'_{k_0}$. 

ii) [Induction step] Think of the case where 
\begin{equation*}
	\textrm{Adjust}'_{k_{i-1}}(x_{i-1})\tilde{\textrm{Query}}_{k_{i}}\preceq \textrm{Query}'_{k'_{i}}
	\tag{\emph{Assumption 1}}
\end{equation*} 
holds in Eq.~(\ref{eq:mode_relation}) for arbitrary $x_0x_1\cdots x_{i-1}$. 
We prove the statement also holds in the case of $x_0x_1\cdots x_{i}$ using a proof by contradiction. 
Let us assume, for some $x_0x_1\cdots x_{i}$, and $\tilde{\textrm{Query}}_{k_{i+1}}\in\tilde{\textsc{Mode}}_{k_{i+1}}$, that 
\begin{equation*}
	\forall \textrm{Query}'_{k'_{i+1}}\in\textsc{Mode}'_{k'_{i+1}}: \textrm{Adjust}'_{k_{i}}(x_{i})\tilde{\textrm{Query}}_{k_{i+1}}\npreceq\textrm{Query}'_{k'_{i+1}} 
	\tag{\emph{Assumption 2}}
\end{equation*} 
Here, we use the notation 
\begin{equation}
	l_{i+1}(a) = \textrm{Point}_{k_i}(a),\:\:l'_{i+1}(a) = \textrm{Point}'_{k'_i}(a)
\end{equation}
for $a\in\mathbb{A}_M$. 
Naturally, $l_{i+1}(x_i)=k_{i+1}$ and $l'_{i+1}(x_i)=k'_{i+1}$.

(\emph{Substep 1} in Fig.~\ref{fig:chart_step_d}) We can write the expanded codeword of $\tilde{T}_{k_i}$ for $x_i$ as 
\begin{eqnarray}
	\tilde{\textrm{Expcw}}_{k_i}(x_i)&\equiv&\textrm{Cword}_{k_i}(x_i)\tilde{\textrm{Query}}_{k_{i+1}}\nonumber\\
	&=&\textrm{Adjust}'_{k_{i-1}}(x_{i-1})\oslash(\textrm{Cword}'_{k'_i}(x_i)\textrm{Adjust}'_{k_i}(x_i)\tilde{\textrm{Query}}_{k_{i+1}})
\end{eqnarray}
using $\tilde{\textrm{Query}}_{k_{i+1}}$ of \emph{Assumption 2}. 
Due to the same assumption and Eq.~(\ref{eq:mode_relation}), any expanded codeword of $T'_{k'_i}$ for $x_i$, 
\begin{equation}
	\textrm{Expcw}'_{k'_i}(x_i)\equiv \textrm{Cword}'_{k'_i}(x_i)\textrm{Query}'_{k'_{i+1}},
\end{equation} 
meets $\textrm{Adjust}'_{k_{i-1}}(x_{i-1})\oslash\textrm{Expcw}'_{k'_i}(x_i)\prec \tilde{\textrm{Expcw}}_{k_i}(x_i)$. 

(\emph{Substep 2} in Fig.~\ref{fig:chart_step_d}) Meanwhile, any expanded codeword of $\tilde{T}_{k_i}$ for $a\neq x_i$ is written as
\begin{equation}
	\tilde{\textrm{Expcw}}_{k_i}(a)\equiv \textrm{Adjust}'_{k_{i-1}}(x_{i-1})\oslash(\textrm{Cword}'_{k'_i}(a)\textrm{Adjust}'_{k_i}(a)\tilde{\textrm{Query}}_{l_{i+1}(a)})
\end{equation}
with $\tilde{\textrm{Query}}_{l_{i+1}(a)}\in \tilde{\textsc{Mode}}_{l_{i+1}(a)}$. 
According to Eq.~(\ref{eq:mode_relation}), there is always some $\textrm{Query}'_{l'_{i+1}(a)}\in \textsc{Mode}'_{l'_{i+1}(a)}$ satisfying $\textrm{Adjust}'_{k_i}(a)\tilde{\textrm{Query}}_{l_{i+1}(a)}\parallel \textrm{Query}'_{l'_{i+1}(a)}$. 
Since we can write the expanded codeword of $\{T'_k\}$ for $a\neq x_i$ as
\begin{equation}
	\textrm{Expcw}'_{k'_i}(a)\equiv \textrm{Cword}'_{k'_i}(a)\textrm{Query}'_{l'_{i+1}(a)},
\end{equation}
we have
\begin{equation}
	\textrm{Adjust}'_{k_{i-1}}(x_{i-1})\oslash\textrm{Expcw}'_{k'_i}(a)\parallel \tilde{\textrm{Expcw}}_{k_i}(a). 
	\label{eq:Ttilde_Tdash_expand_corresp}
\end{equation}

\begin{figure}[tb]
	\begin{center}
		\includegraphics[width=8cm,  bb=0 0 502 241]{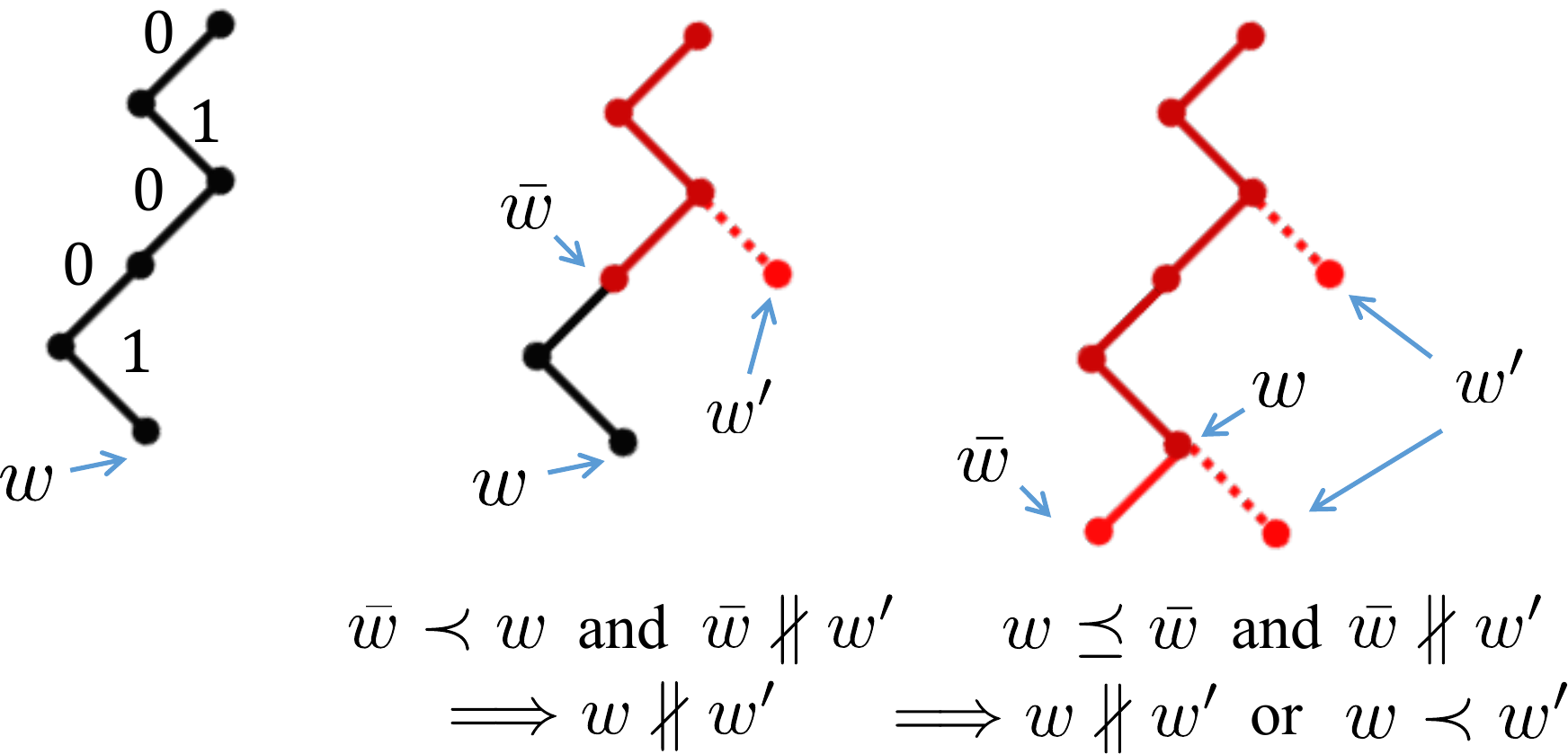}
	\end{center}
	\caption{Example of the relations between binary strings satisfying $w\parallel \bar{w}$ and $\bar{w}\nparallel w'$, which leads to $w'\npreceq w$. }
	\label{fig:three_string_relation_ex}
\end{figure}

Following {\it Rule \ref{rle:decodable}} \ref{srle:prefix_free_expansion} for $T'_{k'_i}$, it must be 
\begin{eqnarray}
	\textrm{Expcw}'_{k'_i}(x_i)&\nparallel& \textrm{Expcw}'_{k'_i}(a) \nonumber\\
	\iff \textrm{Adjust}'_{k_{i-1}}(x_{i-1})\oslash\textrm{Expcw}'_{k'_i}(x_i)&\nparallel& \textrm{Adjust}'_{k_{i-1}}(x_{i-1})\oslash\textrm{Expcw}'_{k'_i}(a).
	\label{eq:rule_3_a_Tdash}
\end{eqnarray}
Combining it with Eq.~(\ref{eq:Ttilde_Tdash_expand_corresp}) gives 
\begin{equation}
	\textrm{Adjust}'_{k_{i-1}}(x_{i-1})\oslash\textrm{Expcw}'_{k'_i}(x_i)\npreceq \tilde{\textrm{Expcw}}_{k_i}(a),  
	\label{eq:Ttilde_Tdash_expand_relation}
\end{equation}
whose derivation uses the idea illustrated in Fig.\ref{fig:three_string_relation_ex}. 
Since $\textrm{Adjust}'_{k_{i-1}}(x_{i-1})\oslash\textrm{Expcw}'_{k'_i}(x_i)\prec \tilde{\textrm{Expcw}}_{k_i}(x_i)$ under \emph{Assumption 2}, it must be $\tilde{\textrm{Expcw}}_{k_i}(a)\nprec \textrm{Adjust}'_{k_{i-1}}(x_{i-1})\oslash\textrm{Expcw}'_{k'_i}(x_i)$ to obey $\tilde{\textrm{Expcw}}_{k_i}(a)\nparallel \tilde{\textrm{Expcw}}_{k_i}(x_i)$ from {\it Rule \ref{rle:decodable}} \ref{srle:prefix_free_expansion} for $\tilde{T}_{k_i}$. 
Therefore, it must be $\textrm{Adjust}'_{k_{i-1}}(x_{i-1})\oslash\textrm{Expcw}'_{k'_i}(x_i)\nparallel \tilde{\textrm{Expcw}}_{k_i}(a)$. 

(\emph{Substep 3} in Fig.~\ref{fig:chart_step_d}) According to the definition of $F_{\rm red}$ and $F_{\rm full}$, $ \tilde{\textrm{Expcw}}_{k_i}(x_i)\in\tilde{\textsc{Expand}}_{k_i}(x_i)=\{\textrm{Cword}_{k_i}(x_i)\tilde{\textrm{Query}}_{k_{i+1}}\mid\tilde{\textrm{Query}}_{k_{i+1}}\in F_{\rm red}(\textsc{Mode}_{k_{i+1}})\}$ implies that
we can make from any ${\textrm{Prefix}}\prec \tilde{\textrm{Expcw}}_{k_i}(x_i)$, by using some ${\textrm{Suffix}}\in\mathbb{W}$, a binary string satisfying $\forall w\in \tilde{\textsc{Expand}}_{k_i}(x_i):{\textrm{Prefix}}\:{\textrm{Suffix}}\nparallel w$. 
Therefore, we can make such ${\textrm{Prefix}}\:{\textrm{Suffix}}\equiv {\textrm{Unused}}$ under the condition of $\textrm{Adjust}'_{k_{i-1}}(x_{i-1})\oslash\textrm{Expcw}'_{k'_i}(x_i)\preceq {\textrm{Unused}}$, awing to  $\textrm{Adjust}'_{k_{i-1}}(x_{i-1})\oslash\textrm{Expcw}'_{k'_i}(x_i) \prec\tilde{\textrm{Expcw}}_{k_i}(x_i)$. 
Because of $\textrm{Adjust}'_{k_{i-1}}(x_{i-1})\oslash\textrm{Expcw}'_{k'_i}(x_i)\nparallel \tilde{\textrm{Expcw}}_{k_i}(a)$, 
it is also $\forall w\in\tilde{\textsc{Expand}}_{k_i}(a):{\textrm{Unused}}\nparallel w$. 
Eventually, under \emph{Assumption 2}, ${\textrm{Unused}}$ satisfies $\forall w\in \tilde{\textsc{Expands}}_{k_i}:\textrm{Unused}\nparallel w$. 

(\emph{Substep 4} in Fig.~\ref{fig:chart_step_d}) We can extend from any prefix of $\tilde{\textrm{Expcw}}_{k_i}(x_i)$ to make $\textrm{Unused}$, which means $F_{\rm full}(\tilde{\textsc{Expands}}_{k_i})$ does not include any prefix of $\tilde{\textrm{Expcw}}_{k_i}(x_i)$. 
Meanwhile, we know $\tilde{\textrm{Expcw}}_{k_i}(x_i)\in F_{\rm full}(\tilde{\textsc{Expands}}_{k_i})$ from the definition of $F_{\rm full}$. 
Therefore, $\tilde{\textrm{Expcw}}_{k_i}(x_i)\in F_{\rm red}(\tilde{\textsc{Expands}}_{k_i})$. 
Since $\{\tilde{T}_k\}$ is a full code-tree set, $\tilde{\textrm{Expcw}}_{k_i}(x_i)\in F_{\rm red}(\tilde{\textsc{Mode}}_{k_i})=\tilde{\textsc{Mode}}_{k_i}$ also holds. 

Combining $\tilde{\textrm{Expcw}}_{k_i}(x_i)\in \tilde{\textsc{Mode}}_{k_i}$ with \emph{Assumption 1}, there must be some string in $\textsc{Mode}'_{k'_i}$ that has $\textrm{Adjust}'_{k_{i-1}}(x_{i-1})\tilde{\textrm{Expcw}}_{k_i}(x_i)=\textrm{Cword}'_{k'_i}(x_i)\textrm{Adjust}'_{k_i}(x_i)\tilde{\textrm{Query}}_{k_{i+1}}$ as its prefix. 
However, such string cannot be a prefix of any expanded codeword in $\textsc{Expand}'_{k'_i}(x_i)$ because of \emph{Assumption 2}.
Moreover, from $\textrm{Adjust}'_{k_{i-1}}(x_{i-1})\oslash\textrm{Expcw}'_{k'_i}(x_i)\prec \tilde{\textrm{Expcw}}_{k_i}(x_i)$ and Eq.~(\ref{eq:rule_3_a_Tdash}), it has to be $\textrm{Adjust}'_{k_{i-1}}(x_{i-1})\tilde{\textrm{Expcw}}_{k_i}(x_i)\nparallel\textrm{Expcw}'_{k'_i}(a)$. 
These facts suggest that $\textsc{Mode}'_{k'_{i+1}}$ must have some string that cannot be a prefix of any expanded codeword in $\textsc{Expands}'_{k'_i}$, which conflicts with Eq.~(\ref{eq:assum_no_invalid_query_Tdash}). 
Therefore, $\textrm{Adjust}'_{k_{i}}(x_{i})\tilde{\textrm{Query}}_{k_{i+1}}\preceq \textrm{Query}'_{k'_{i+1}}$ also holds. \\

As a result, we have
\begin{equation}
	\forall \tilde{\textrm{Query}}_{k_{i}}\in\tilde{\textsc{Mode}}_{k_{i}}, \exists \textrm{Query}'_{k'_i}\in\textsc{Mode}'_{k'_i}: \textrm{Adjust}'_{k_{i-1}}(x_{i-1})\tilde{\textrm{Query}}_{k_{i}}\preceq \textrm{Query}'_{k'_{i}}, 
	\label{eq:mode_relation_real}
\end{equation}
and therefore
\begin{equation}
	\forall \tilde{\textrm{Query}}_{k_{i}} \in \tilde{\textsc{Mode}}_{k_{i}}, \exists \textrm{Query}'_{k'_i}\in \textsc{Mode}'_{k'_i}: \|\tilde{\textrm{Query}}_{k_{i}}\|_{\rm len}\le \|\textrm{Query}'_{k'_i}\|_{\rm len} 
\end{equation}
for any $i\in\mathbb{Z}^+$. 
From the result of step b, the decoding delay of a code using $\{\tilde{T}_k\}$ cannot be longer than the one using any $\{T'_k\}$. 
$\qquad \blacksquare$

\subsection{Representation of modes by fixed-length strings}
Owing to the above theorem, we can ignore useless codes, whose delay we can shorten without changing the codewords. 
We need only to consider the basic modes, which have no common prefix except `$\lambda$' and are invariable by $F_{\rm red}$. 
It should be noted that the basic mode corresponding to modes with 1 bit always be $\{$`$\lambda$'$\}$. 
For this reason, 1 bit of decoding delay never contributes to the compression efficiency in binary codes, which is consistent with the fact proven in the previous work \cite{ref:huf_opt_aifv1}. 

For decoding delay longer than 1 bit, the basic modes can be represented simpler using their constraints. 
To introduce such representation, we add a notation: 
\begin{itemize}
	\item $B_N$: The member of $\mathbb{M}$ including all the binary strings of length $N$. 
\end{itemize}
All of the basic modes can be written by this notation as follows. 
\mytheory{thm}{Basic mode variation}{
	\label{thm:bm_var}
	For an arbitrary $\textsc{Mode}\in\mathbb{M}$, suppose $\tilde{\textsc{Mode}}=\{f_{\rm cmn}(\textsc{Mode})\oslash\hat{\textrm{Query}}\mid \hat{\textrm{Query}}\in F_{\rm red}(\textsc{Mode})\}$. If $\forall\tilde{\textrm{Query}}\in\tilde{\textsc{Mode}}:\|\tilde{\textrm{Query}}\|_{\rm len} \le N$ ($N\ge 2$), 
	\begin{equation}
		\exists \textsc{Lb},\textsc{Ub}\subset B_{N-1}: F_{\rm red}(\{\text{`0'}\textrm{Lbin}\mid \textrm{Lbin}\in \textsc{Lb}\}\cup \{\text{`1'}\textrm{Ubin}\mid \textrm{Ubin}\in \textsc{Ub}\})=\tilde{\textsc{Mode}}. 
\end{equation} }
\underline{{\it Proof}}: 
It is obvious from the definition that, for any $n\in\mathbb{N}$, $\textsc{Words}\in\mathbb{M}$ and $w\in \textsc{Words}$, 
\begin{equation}
	F_{\rm red}((\textsc{Words}\setminus\{w\})\cup \{wb\mid b\in B_n\}\})=F_{\rm red}(\textsc{Words}). 
\end{equation}Using {\it Lemma \ref{lmm:redset}} \ref{slmm:preserve_cmn_prefix} and \ref{slmm:irreducible}, $F_{\rm red}(\tilde{\textsc{Mode}}) = \tilde{\textsc{Mode}}$. 
Therefore, $\tilde{\textsc{Mode}}$ can be rewritten as
\begin{equation}
	\tilde{\textsc{Mode}} = F_{\rm red}(\tilde{\textsc{Mode}}) = F_{\rm red}(\{\tilde{\textrm{Query}}b_{N-\|\tilde{\textrm{Query}}\|_{\rm len}}\mid b_{N-\|\tilde{\textrm{Query}}\|_{\rm len}}\in B_{N-\|\tilde{\textrm{Query}}\|_{\rm len}}, \tilde{\textrm{Query}}\in \tilde{\textsc{Mode}}\}), 
\end{equation}
and every binary string in the set $\{\tilde{\textrm{Query}}b_{N-\|\tilde{\textrm{Query}}\|_{\rm len}}\}$ is $N$-length. 

Additionally, $f_{\rm cmn}(\tilde{\textsc{Mode}}) = f_{\rm cmn}(\textsc{Mode})\oslash f_{\rm cmn}(\textsc{Mode})=\text{`$\lambda$'}$ due to {\it Lemma \ref{lmm:redset}} \ref{slmm:preserve_cmn_prefix}. 
Thus, $\tilde{\textsc{Mode}}$ must contain some strings respectively starting with `0' and `1'. 
$\qquad \blacksquare$\\\\

Owing to the above theorem, we only have to consider the combination of $N$-bit strings when setting the modes for $N$-bit-delay AIFV codes. 
For example, $2$-bit-delay AIFV codes have nine patterns of basic modes, which have no common prefix except `$\lambda$' and are invariable by $F_{\rm red}$: 
$\{$`$\lambda$'$\}$, $\{$`0', `10'$\}$, $\{$`0', `11'$\}$, $\{$`00', `1'$\}$, $\{$`00', `10'$\}$, $\{$`00', `11'$\}$, $\{$`01', `1'$\}$, $\{$`01', `10'$\}$, and $\{$`01', `11'$\}$. 
They can be rewritten as some sets of 2-bit binary strings by altering as `$\lambda$' $\to$ $\{$`00', `01', `10', `11'$\}$, `0' $\to$ $\{$`00', `01'$\}$ and `1' $\to$ $\{$`10', `11'$\}$. 
The rewritten basic modes always include 2-bit binary strings beginning with `0' and ones with `1'. 
Therefore, they can be represented as a pair of 1-bit binary string sets $\textsc{Lb}$ and $\textsc{Ub}$: 
When $\tilde{\textsc{Mode}}=\{$`$\lambda$'$\}$, $\textsc{Lb}=\textsc{Ub}=\{$`0', `1'$\}$; 
when $\tilde{\textsc{Mode}}=\{$`0', `10'$\}$, $\textsc{Lb}=\{$`0', `1'$\}$ and $\textsc{Ub}=\{$`0'$\}$.  

\subsection{Representation of decodable condition by intervals}
\begin{figure}[tb]
	\begin{center}
		\includegraphics[width=5cm,  bb=0 0 324 349]{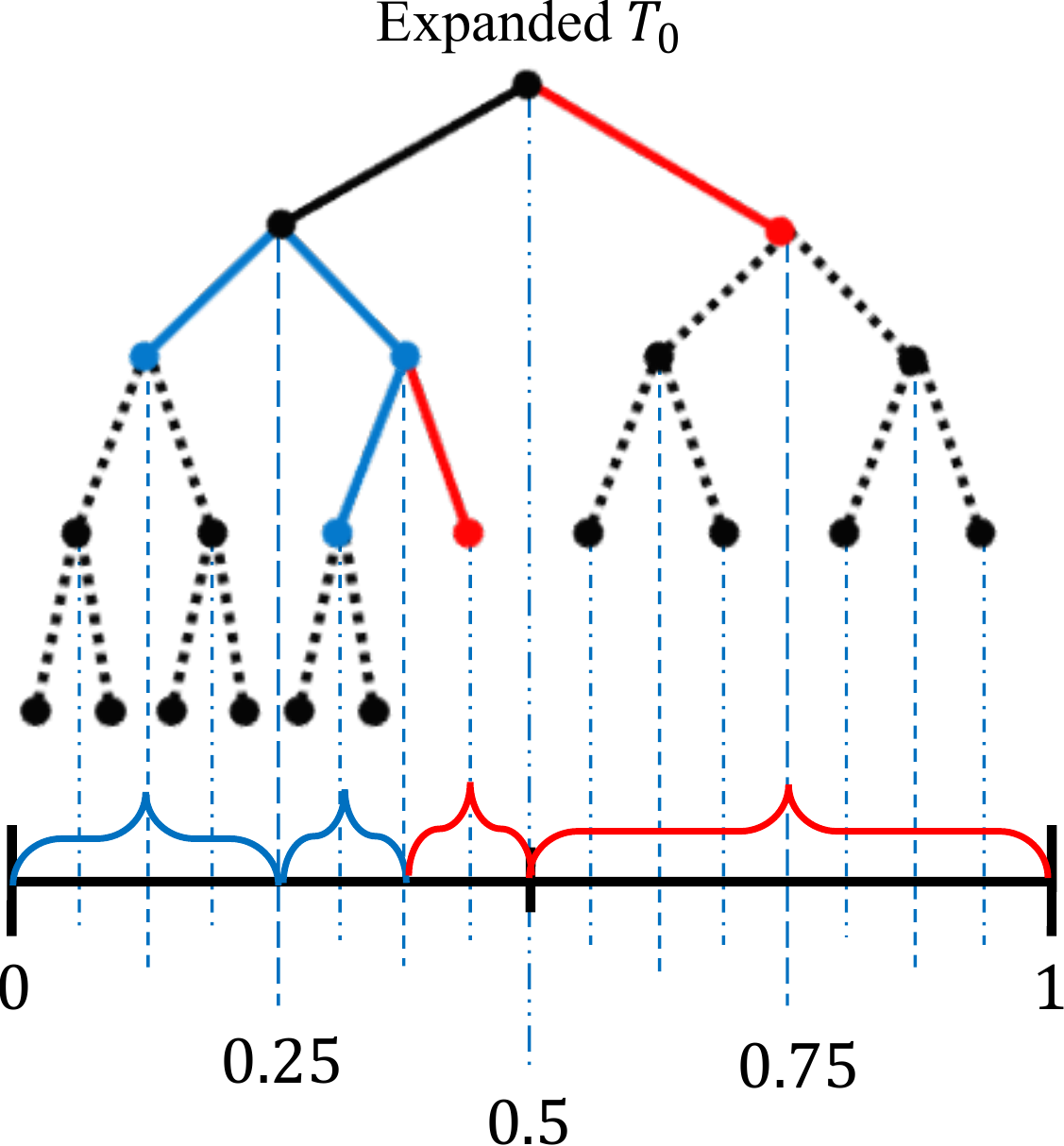}
	\end{center}
	\caption{Correspondence of nodes and probability intervals of the expanded $T_0$ in Fig.~\ref{fig:expand_ex}. }
	\label{fig:prob_interval_ex}
\end{figure}
Since every basic mode is a member of $\mathbb{PF}$, {\it Rule \ref{rle:decodable}} \ref{srle:prefix_free_expansion} can be equivalently written in a simpler way when the code trees contain only the basic modes: 
\mytheory{rle}{Equivalent {\it Rule \ref{rle:decodable}} {\rm a} for $\{T_k\}$ with the basic modes}
{
	\label{rle:eq_decodable_1}
	\begin{enumerate}
		\item $\forall k: \textsc{Expands}_k\in\mathbb{PF}$. 
	\end{enumerate} 
} 
In other words, we must keep the expanded codeword sets to satisfy the prefix condition if we want to design the proposed AIFV codes. 
It is well-known that Kraft's inequality allows us to check whether there are some codes satisfying the prefix condition. 
However, we cannot know from the inequality the expanded codeword set is following {\it Rule \ref{rle:decodable}}. 
We here show one way to tackle this problem by mapping the codewords to intervals among the real number line, 
which is a similar approach to the range coding \cite{ref:hbdc}. 
Let us use notations and functions as below. 
\begin{itemize}
	\item $\myinter{x_l, x_u}$: $\{x\in\mathbb{R}\mid x_l\le x<x_u\}$, an interval between $x_l$ and $x_u$ ($\in \mathbb{R}$). 
	\item $\mathbb{PI}$: $\{R\subset \myinter{0, 1}\}$, a set of all probability intervals, intervals included between $0$ and $1$. 
	\item $f_{\rm dec}$: $\mathbb{W}\to \myinter{0, 1}$. $f_{\rm dec}($`$y_0y_1\cdots y_{L-1}$'$)=\sum_{i\in \mathbb{Z}^+_{<L}}y_i/2^{i+1}$ for $y_i\in\{0,1\}$.  
	\item $F_{\rm prob}$: $\mathbb{W}\to \mathbb{PI}$. $F_{\rm prob}(w)=\myinter{f_{\rm dec}(w), f_{\rm dec}(w)+1/2^{\|w\|_{\rm len}}}$. 
\end{itemize}
Since 
\begin{equation}
	F_{\rm prob}(\text{`$y_0y_1\cdots y_{L-1}$'})=\myInter{f_{\rm dec}(\text{`$y_0y_1\cdots y_{L-1}$'}),\:\:f_{\rm dec}(\text{`$y_0y_1\cdots y_{L-1}$'})+\frac{1}{2^L}},
\end{equation}
the boundaries of the interval given by $F_{\rm prob}($`$y_0y_1\cdots y_{L-1}$'$)$ can be written in binary numbers as $0.y_0y_1\cdots y_{L-1}00\cdots$ and $0.y_0y_1\cdots y_{L-1}11\cdots$. 
Therefore, for $w_1,w_2\in \mathbb{W}$, $F_{\rm prob}(w_1)\cap F_{\rm prob}(w_2)$ is not empty if and only if $w_1\parallel w_2$, and we get an alternative condition for checking {\it Rule \ref{rle:decodable}} \ref{srle:prefix_free_expansion} ({\it Rule \ref{rle:eq_decodable_1}}): 
\begin{equation}
	\forall \textsc{Expands}\in \mathbb{M}: \textsc{Expands}\in\mathbb{PF} \iff \bigcup_{w_1,w_2\in \textsc{Expands}, w_1\neq w_2} (F_{\rm prob}(w_1)\cap F_{\rm prob}(w_2))=\emptyset.
	\label{eq:pi_rule3_a}
\end{equation} 
Fig.~\ref{fig:prob_interval_ex} depicts an example of the tree representing the expanded codewords of $T_0$ in Fig.~\ref{fig:expand_ex}. 
The modes of the next code trees corresponding to $a$ and $b$ are respectively $F_{\rm red}(\{$`011', `100', `101', `110', `111'$\})$ and $F_{\rm red}(\{$`000', `001', `010', `011', `100', `101'$\})$, 
and the probability intervals of the expanded codewords $\myinter{0, 0.25}$, $\myinter{0.25, 0.375}$, $\myinter{0.375, 0.5}$, and $\myinter{0.5, 1}$ do not overlap with each other. 

It is also obvious that for $w_1,w_2\in \mathbb{W}$, $F_{\rm prob}(w_1)\subseteq F_{\rm prob}(w_2)$ if and only if $w_2\preceq w_1$, and 
\begin{eqnarray}
	\forall (\textsc{Mode}\in \mathbb{PF}, \textsc{Expands}\in \mathbb{M}, \textrm{Expcw} \in \textsc{Expands}):&&\nonumber\\
	(\exists \textrm{Query} \in \textsc{Mode}: \textrm{Query} \preceq \textrm{Expcw}) &\iff& F_{\rm prob}(\textrm{Expcw}) \subseteq \bigcup_{\textrm{Query} \in \textsc{Mode}} F_{\rm prob}(\textrm{Query}).
	\label{eq:pi_rule3_b}
\end{eqnarray} 
This condition can be used to check {\it Rule \ref{rle:decodable}} \ref{srle:prefix_in_mode}. 
Especially in the case of a full code-tree set, the right-hand side of Eq.~(\ref{eq:pi_rule3_b}) will be equal. 

Owing to Eqs.~(\ref{eq:pi_rule3_a}) and (\ref{eq:pi_rule3_b}), we can check the decodability only by placing the intervals on the real number line. 
This property is expected to be useful in constructing the proposed codes. 

%%%%%%%%%%%%%%%%%%%%%%%%%%%%%%%%%%%%%%%%%%%%%%%%%%%%%%%%%%%%%%%%%%%%%%%%%%%%%%%%
\section{Further discussions}
\label{sec:disscuss}
\subsection{Relationship with conventional codes}
\begin{figure}[!tb]
	\begin{center}
		\subfigure[Optimal AIFV-$3$ code introduced in \cite{ref:mcopt_proof}. ]{
			\includegraphics[width=7.1cm,  bb=0 0 588 428]{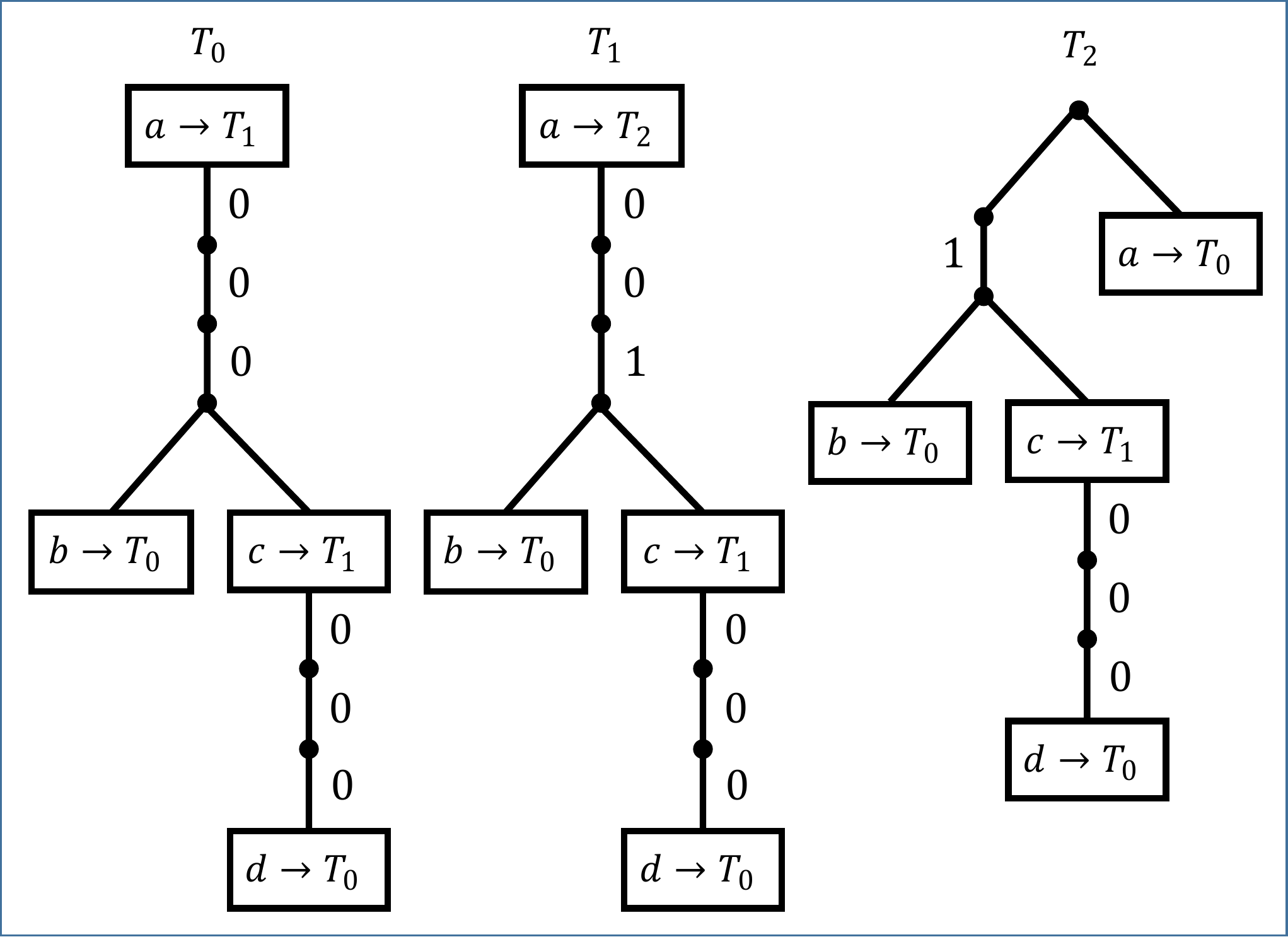}}
		\subfigure[Proposed 3-bit-delay AIFV code. ]{
			\includegraphics[width=9.4cm,  bb=0 0 776 428]{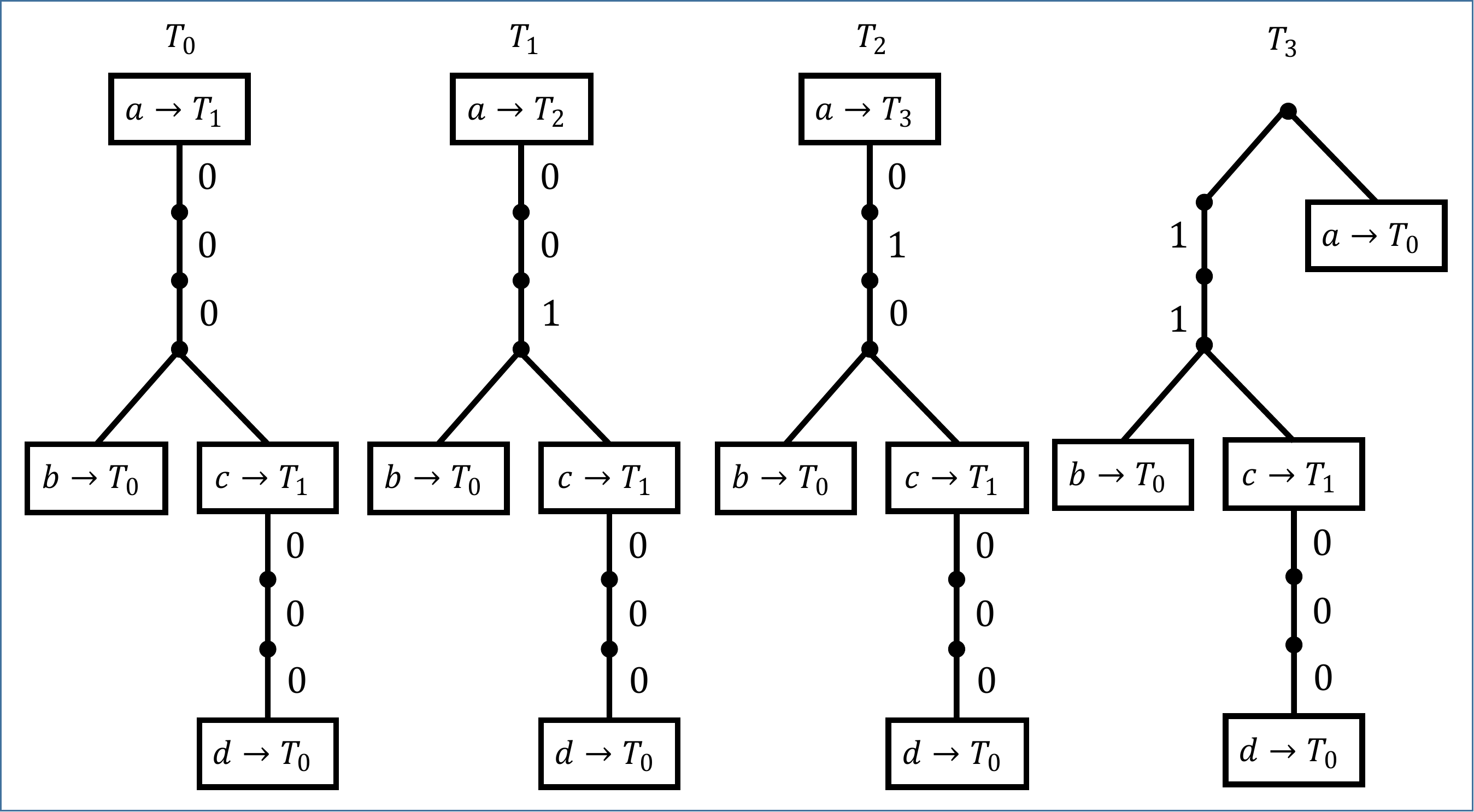}}
	\end{center}
	\caption{Example of codes for $\mathbb{A}_4$ distributed as $p_{\rm src}(a)=0.9$, $p_{\rm src}(b)=0.05$, $p_{\rm src}(c)=0.049$, and $p_{\rm src}(d)=0.001$. The proposed 3-bit-delay AIFV code achieves a shorter expected code length than the AIFV-$3$ one. }
	\label{fig:opt_aifv3_ex}
\end{figure}
The conventional AIFV-$m$ codes can be interpreted, when $m=N$, as $N$-bit-delay AIFV codes with their code trees $\{T_k\}\in\mathbb{DT}_M$ limited to 
\begin{equation}
	\bigcup_{\textrm{Query}_k \in \textsc{Mode}_k} F_{\rm prob}(\textrm{Query}_k)=\myInter{0, 1}\:\:\text{or}\:\:\myInter{2^{-(N-n+1)}, 1}, 
\end{equation} 
with $n=1,2,\cdots,N-1$. 
Fig.~\ref{fig:opt_aifv3_ex} shows an example of the difference between the conventional AIFV-$m$ and proposed $N$-bit-delay AIFV codes. 
The AIFV-$3$ code is the one introduced in the previous work \cite{ref:mcopt_proof} as optimal for $\mathbb{A}_4$ whose probabilities are $p_{\rm src}(a)=0.9$, $p_{\rm src}(b)=0.05$, $p_{\rm src}(c)=0.049$, and $p_{\rm src}(d)=0.001$. 
We can see that Fig.~\ref{fig:opt_aifv3_ex} (a) satisfies {\it Rule \ref{rle:aifvm}} for AIFV-$m$ codes, which is equivalent to using only the modes $F_{\rm red}(\{$`000', `001', `010', `011', `100', `101', `110', `111'$\})$, $F_{\rm red}(\{$`001', `010', `011', `100', `101', `110', `111'$\})$, and $F_{\rm red}(\{$`010', `011', `100', `101', `110', `111'$\})$. 
The expected code length of the AIFV-3 code is 
\begin{eqnarray}
	&&(0\times p_{\rm src}(a) + 4\times p_{\rm src}(b) + 4\times p_{\rm src}(c) + 7\times p_{\rm src}(d))\times \Pi_0\nonumber\\
	&+& (0\times p_{\rm src}(a) + 4\times p_{\rm src}(b) + 4\times p_{\rm src}(c) + 7\times p_{\rm src}(d))\times \Pi_1\nonumber\\
	&+& (1\times p_{\rm src}(a) + 3\times p_{\rm src}(b) + 3\times p_{\rm src}(c) + 6\times p_{\rm src}(d))\times \Pi_2 =0.655\cdots
\end{eqnarray} 
where the stationary probability $\Pi_k$ for $T_k$ is given by 
\begin{equation}
	(\Pi_0, \Pi_1, \Pi_2) = (\Pi_0, \Pi_1, \Pi_2)\left(
	\begin{array}{ccc}
		p_{\rm src}(b)+p_{\rm src}(d) & p_{\rm src}(a)+p_{\rm src}(c) & 0\\
		p_{\rm src}(b)+p_{\rm src}(d) & p_{\rm src}(c) & p_{\rm src}(a)\\
		p_{\rm src}(a)+p_{\rm src}(b)+p_{\rm src}(d) & p_{\rm src}(c) & 0
	\end{array}
	\right).
\end{equation} 
The expected code length of the $N$-bit-delay AIFV code is similarly calculated as $0.604\cdots$. 
It is much closer to the entropy $0.576\cdots$ than AIFV-3 code. 
Although the proposed code uses more code trees, it still keeps the decoding delay within 3 bits and makes more use of the allowed delay. 

As stated above, the conventional AIFV-$m$ codes always use modes whose intervals are continuous. 
This is also true for arithmetic coding if we interpret it as AIFV codes using code-tree sets. 
The relationship between arithmetic coding and AIFV codes has been reported in previous works \cite{ref:aifv_arith1,ref:aifv_arith2}. 
Arithmetic coding is identical to the one of the proposed codes when $N\to\infty$, using infinite code trees with the intervals of their modes constrained to be continuous. 

Some class of AIFV codes reported \cite{ref:aifv_treepair1,ref:aifv_treepair2} uses modes whose intervals can be discontinuous. 
It is a subclass of the proposed codes which limits the code trees to two using every available mode.

\subsection{Open questions}
One of the essential questions remaining is the theoretical worst-case redundancy of the proposed codes. 
From a very conservative perspective, it is lower than or equal to the redundancy stated in the previous work \cite{ref:aifv2}. 
However, the results are based on the condition where the number of code trees is identical to the decoding delay, which is not a very reasonable assumption in this case. 
There must be a stricter bound to evaluate the redundancy. 
Recently, some properties have been found for general codes decodable within finite lengths of decoding delay \cite{ref:kdelay_prop,ref:aifv2_opt_proof}. 
These results may be combined with the proposed theories. 

Another interesting question is how to obtain optimal codes for a given source. 
Although we presented a method making some $N$-bit-delay AIFV code from a given VV code in {\it Proof of Theorem \ref{thm:generality}}, 
its main purpose was to show the existence of some code-tree set corresponding to the given code. 
It requires setting a new code tree $T_{k_s}$ every time we break down the given code and may be impractical for constructing general VV codes. 
One possible approach for practical construction is to divide the code-tree optimization problem into tree-wise forms, as the previous works \cite{ref:mcopt_proof,ref:mcopt} do. 
The decodable condition introduced in the paper will be helpful in designing the optimization algorithm in such a case. 
Since the proposed code is a wide class, including any conventional codes presented here, 
it is expected to outperform other codes if we have such an algorithm. 

%%%%%%%%%%%%%%%%%%%%%%%%%%%%%%%%%%%%%%%%%%%%%%%%%%%%%%%%%%%%%%%%%%%%%%%%%%%%%%%%
\section{Conclusions}
\label{sec:conclusions}
We presented $N$-bit-delay AIFV codes, which can represent every code we can make when permitting decoding delay up to $N$ bits. 
By introducing the concepts of modes and expanded codewords, we explained the relationships between the decoding delay and the code structure. 
It was shown that to construct uniquely decodable codes, the expanded codewords should be prefix-free and have prefix included in the corresponding mode. 
Additionally, the decoding delay of the proposed code was shown to be the maximum-length string in the modes. 

Then, we detected the class of modes, basic modes, that achieves minimum decoding delay among the codes giving the same codeword sequences. 
The idea of basic modes greatly reduced the freedom of modes we have to consider. 
Moreover, we derived a reasonable formulation of constraints for decodability in the case of basic modes. 
Based on the conversion of binary strings to intervals in the real number line, it was shown that we need only to compare the intervals corresponding to expanded codewords and modes. 
This formulation will make it easier to guarantee decodability numerically when constructing codes. 

Although there are still many questions, the theoretical results presented in this paper must be essential for future study.

\bibliographystyle{ieeetr}

\end{document}